\documentclass[aip,jcp,preprint,superscriptaddress]{revtex4}

\usepackage[english]{babel}
\usepackage{graphicx} 
\usepackage{amsmath,amssymb,amsfonts} 
\usepackage{color,soul}
\usepackage{epstopdf}
\usepackage{array}
\usepackage{multirow}
\usepackage{ulem}
\usepackage{bm}

\newcommand{\ea}{\textit{et~al.}}
\newcommand{\mrm}{\mathrm}
\newcommand{\dd}{\mathrm{d}}
\newcommand{\mexp}{\mathrm{e}}
\newcommand{\erf}{\mathrm{erf}}
\newcommand{\zbar}{\bar{z}}
\newcommand{\mT}{\mathcal{T}}
\newcommand{\mQ}{\mathcal{Q}}
\newcommand{\mV}{\mathcal{V}}

\begin{document}

\bibliographystyle{apsrev4-1}

\title{Affinity, kinetics, and pathways of anisotropic ligands binding to hydrophobic model pockets} 

\author{R. Gregor Wei{\ss}}
\thanks{gregor.weiss@physik.hu-berlin.de} 
\affiliation{Institut f{\"u}r Physik, Humboldt-Universit{\"a}t zu Berlin, Newtonstrasse~15, D-12489 Berlin, Germany}
\affiliation{Laborartory of Physical Chemistry, ETH Z{\"u}rich, CH-8093 Z{\"u}rich, Switzerland}
\author{Richard Chudoba}
\affiliation{Institut f{\"u}r Physik, Humboldt-Universit{\"a}t zu Berlin, Newtonstrasse~15, D-12489 Berlin, Germany}
\affiliation{Research Group Simulations of Energy Materials, Helmholtz-Zentrum Berlin, Hahn-Meitner-Platz 1, D-14109 Berlin, Germany}
\affiliation{Physikalisches Institut, Albert-Ludwigs-Universit{\"a}t Freiburg, Hermann-Herder Strasse~3, D-79104 Freiburg, Germany}
\author{Piotr Setny}
\affiliation{Centre of New Technologies, University of Warsaw, Stefana Banacha 2c, 02-927 Warsaw, Poland}
\author{Joachim Dzubiella}
\thanks{joachim.dzubiella@helmholtz-berlin.de}
\affiliation{Institut f{\"u}r Physik, Humboldt-Universit{\"a}t zu Berlin, Newtonstrasse~15, D-12489 Berlin, Germany}
\affiliation{Research Group Simulations of Energy Materials, Helmholtz-Zentrum Berlin, Hahn-Meitner-Platz 1, D-14109 Berlin, Germany}
\affiliation{Physikalisches Institut, Albert-Ludwigs-Universit{\"a}t Freiburg, Hermann-Herder Strasse~3, D-79104 Freiburg, Germany}

\begin{abstract}
Using explicit-water molecular dynamics (MD) simulations of a generic pocket-ligand model we
investigate how chemical and shape anisotropy of small ligands influences  the affinities,
kinetic rates and pathways for their association to hydrophobic binding sites. In particular, we
investigate aromatic compounds, all of similar molecular size, but distinct by various
hydrophilic or hydrophobic residues. We demonstrate that the most hydrophobic sections are in
general desolvated primarily upon binding to the cavity, suggesting that specific hydration of
the different chemical units can steer the orientation pathways via a `hydrophobic torque'.
Moreover, we find that ligands with bimodal orientation fluctuations have significantly
increased kinetic barriers for binding compared to the kinetic barriers previously observed for
spherical ligands due to translational fluctuations. We exemplify that these kinetic barriers,
which are ligand specific, impact both binding and unbinding times for which we observe
considerable differences between our studied ligands.
\end{abstract}

\maketitle

Molecular recognition in aqueous solution is of fundamental importance in living and chemically
engineered systems. As an example, enzymes bind a substrate to an often complementary, concavely
shaped binding site~\cite{enzyme1}. Also, receptors are activated or inhibited if their binding
pockets take up a small molecule, such as a neurotransmitter~\cite{dopamine}, a
hormone~\cite{hormone1} or a pharmaceutical drug~\cite{cb1}. Moreover, this binding principle
from nature is copied in chemical engineering of supramolecular chemistry, where so-called
cavitands~\cite{Gibb:OrgLett, Ashbaugh&Gibb} or macrocycles~\cite{Xue&Huang} are designed as
molecular containers. The superior principle is often pictured by a binding agent  representing
a 'key' or 'guest'  selectively fitting into a complementary shaped 'lock', or 'host',
respectively, giving the names of key-lock or host-guest principles~\cite{key-lock1,
host-guest1}. Modern drug screening and design is based on this principle.

Key-lock binding in water exhibits strong solvent-mediated effects which in fact can be
diverse~\cite{Ball:ChemRev} but are very often of hydrophobic nature. By today, it is well
accepted from computer simulations and experimental studies that hydrophobic protein pockets
comprise strong contributions to the binding affinity of ligands through non-trivial dehydration
effects~\cite{Berne&Friesner, Friesner&Berne, Nair, Rossky, Setny:PRL, Setny:JCTC:2010,
Setny:JACS}. Motivated by the recent recognition of the importance of drug-receptor binding
rates  for the efficacy of the drug~\cite{Pan&Shaw}, most novel studies have now added on the
role of water in hydrophobic association kinetics. In particular, Setny~\ea~\cite{Setny:PNAS}
documented a direct coupling between water fluctuations in hydrophobic pockets and the ligand
binding rate. Using explicit-water MD simulations, they studied the binding of a spherical
ligand to a hydrophobic pocket represented by a hemispherical surface recess in a model
wall~\cite{Setny2006, Setny2007, Setny2008}. They demonstrated that hydrophobically driven
wet-dry fluctuations inside the pocket could lead to locally enhanced ligand friction and
kinetic barriers in the vicinity of the binding site. These findings were consistent with
observations of Berne and coworkers in a study on a similar hydrophobic key-lock model
setup~\cite{Mondal:PNAS}. In a most recent work of ours~\cite{Weiss:JCTC}, we also showed that
increased hydrophobicity by modulating shape or water affinity of the pocket influences the
friction peak and can speed up the binding kinetics.

All previous modeling work on hydrophobic pocket-ligand binding exclusively focused on simple
spherical ligands, mimicking a methane-like molecule~\cite{Setny:PRL, Setny:JCTC:2010,
Setny:JACS, Setny:PNAS, Setny2006, Setny2007, Setny2008, Weiss:JCTC} or other idealized
carbon-based assemblies~\cite{Mondal:PNAS, Tiwary:PNAS, Tiwary:JCP}. Chemical and shape
anisotropy of ligands, however, are of fundamental importance for biological
function~\cite{Ludwig2007, Wittmann2017}. For instance, analyses of a series of chemical
derivatives demonstrated that altering the ensemble of ligand binding orientations changes
signaling output, providing a novel mechanism for titrating allosteric signaling
activity~\cite{Bruning2010}. Other work found that only the orientation of the substrates
correlated with the conjugation capacity in {\it in-vitro} experiments, where the conjugation
reaction proceeded only when the hydroxyl group of the ligand is oriented towards the
coenzyme~\cite{Takaoka2010}. Hence,  the widely used spherical models of ligands is most of the
time inadequate for the identification of potential binding pockets in computational
methods~\cite{Benkaidali2013}. A more systematic investigation of the effects of chemical and
shape anisotropy of a ligand to a hydrophobic pocket on affinity as well as kinetics is
therefore of high fundamental interest. Since complex anisotropic ligands have more degrees of
freedom than a simple sphere, interesting behavior in the coupling to hydration and in the
association pathways can be expected, possibly opening more opportunites in drug design.

So far, fundamental studies on solvent-influenced binding pathways of anisotropic substrates can
be only found in more coarse-grained descriptions discussing the role of solvent depletion in
molecular pair association processes. Kinoshita~\cite{Kinoshito}, for instance, calculated the
depletion potential of two freely rotating plates which favors an association pathway with very
tilted orientations to each other. Moreover, Roth~\ea~\cite{Roth} estimated the association
potential of a rod associating to a planar wall. They found that solvent depletion generates a
torque that will favor an equivalently tilted association pathway. Also in the application to
key-lock models~\cite{König} with a spheroidal ligand the depletion forces were calculated and
found to exhibit comparably high barriers if the ligand faces the pocket with its extended side
during the association process. In essence, they conclude that an aspherical ligand will
non-parallely associate to a concave binding site and only in the last step fold/lay down into
the pocket. Dlugosz~\ea~\cite{Dlugosz} studied the binding time in Brownian simulations of an
spheroidal ligand that electrostatically associated to a concave pocket. If the hydrodynamic
interactions were turned off, the mean binding time was faster than in calculations which
considered hydrodynamic interactions. Hence the water-mediation, i.e., hydrodynamics, plus the
reorientation process of the aspherical ligand increased the absolute binding time.

In this paper, we investigate how chemical and shape anisotropy of small ligands
influence the affinities, kinetic rates and pathways for the association to hydrophobic
binding sites using explicit-water MD simulations of the generic
pocket model used previously~\cite{Setny:PRL, Setny:JCTC:2010, Setny:JACS, Setny:PNAS,
Setny2006, Setny2007, Setny2008, Weiss:JCTC}. We highlight in particular how water
influences the binding and the unbinding of various aromatic ligands to hydrophobic
binding sites, and how the new rotational degrees of freedom couple to
our previously reported water fluctuation effects~\cite{Setny:PNAS, Weiss:JPCB,
Weiss:JCTC}. First, we focus on the binding of benzene to a hydrophobic planar wall
compared to a hydrophobic binding pocket and present the ligand reorientation potential,
pathway, and friction profile. We elaborate on an interpretation of enhanced
friction as kinetic barrier in a rescaled energy landscape which was originally
introduced by Hinczewski~\textit{et al.}~\cite{Hinczewski&Netz} in the context of
protein folding. Moreover, we discuss how the kinetic barrier varies upon our various
ligands, which are all of similar molecular size, but distinct by various hydrophilic or
hydrophobic residues. We describe that the general concept of a
kinetic barrier is not only influenced by the previously reported water fluctuations but
also by the new rotational degree of freedom of the ligands. We analytically discuss
the impact of the resulting slowdown and see that the most significant implications must
follow for the unbinding process, whereas the unbinding time can be influenced by
hundreds of microseconds. Thus the kinetic
barrier can be steered by ligand shape and, in turn, steer the binding
and unbinding rates such that the action of a toxin and the efficacy of a drug can be
optimized.

\section{Methods}

\begin{figure}[t]
\centering \includegraphics[width=8.6cm]{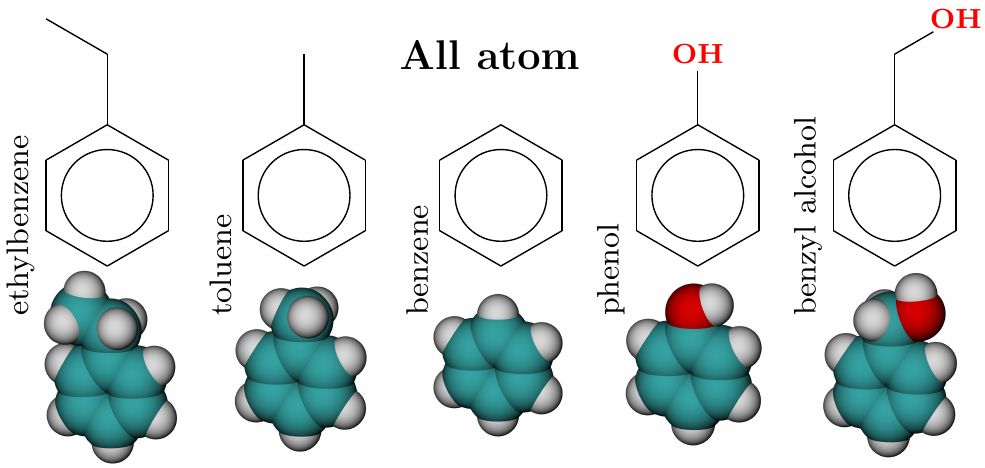}
\caption{All ligands studied in this work contain an aromatic ring. The reference benzene ring
in the center is alkylated with a methyl and ethyl group, respectively, while stepping to the
left creating toluene and ethylbenzene. Stepping to the right, a hydroxyl group and a
hydroxymethyl moiety are respectively introduced to form phenol and benzyl alcohol.}
\label{fig1}
\end{figure}

\subsection{Varyious non-spherical ligands}

In contrast to our previous study~\cite{Weiss:JCTC} in which we modified the physicochemical
properties of the binding site, we investigate here the binding of various aromatic compounds to
hydrophobic binding sites. Therefore we use ethylbenzene, toluene, benzene, phenol and benzyl
alcohol, which are illustrated in Fig.~\ref{fig1}. The aromatic compounds are represented by the
OPLS-AA force field~\cite{Jorgensen1996OPLS-AA} whereas we employ the LINCS algorithm to
constrain all bond lengths. Phenol can be considered to be the most hydrophilic compound based
on its ratio of polar to non-polar solvent-accessible surface area while hydrophobicity
increases for the ligands left and right from it in Fig.~\ref{fig1}. We use the TIP4P model for
water~\cite{Jorgensen1983TIP4P, Jorgensen1985TIP4P}.

\subsection{Constrained simulations for PMF and ligand orientation}

The simulation setup is illustrated in Fig.~2 and 3 which is the same as in our previous
study~\cite{Weiss:JCTC}. The ligand binding process is constrained to one-dimensional diffusion
along $z$, the distance of the ligand center-of-mass perpendicular to the pocketed wall shown in
Fig.~2 and 3. Movement of the center of mass (COM) along $x$- and $y$-direction was strongly
restrained with a harmonic potential with spring constant
$k_{x/y}=42000~\text{kJ~mol}^{-1}~\text{nm}^{-2}$. To probe selected observables as functions of
the ligand separation to the binding site we utilized umbrella sampling simulations along $z$.
In each umbrella setup, the center of mass of the ligand was constrained to a given position
$z_i$ using an external harmonic potential with spring constant
$K=835~\text{kJ~mol}^{-1}~\text{nm}^{-2}$. We define the origin $z=0$ by the first layer of the
wall that is in contact with the water (Fig.~2) or the pocket's bottom, namely the inner crystal
layer of the pocket (Fig.~3). The first umbrella potential was placed at $z=3.14~$\AA. Up to 48
additional umbrella windows with 0.5~\AA~spacing were introduced to cover increasing
ligand-pocket distances. The individual umbrella simulations produced 10~ns with a step size of
2~fs, while the ligand coordinate was stored for every time step and the water coordinates were
stored every 20~fs. The potential of mean force (PMF) was obtained by the weighted histogram
analysis method (WHAM)~\cite{WHAM, HummerWHAM}.

The molecular snapshots in Figs.~\ref{fig2} and \ref{fig3} also illustrate how we define the
ligand orientation by the angle $\theta$. It is the angle between the normal vector of the
aromatic ring and the $z$-axis, and consequentially the angle between the ring's plane spanned
by its ring atoms and the $x$-$y$-plane (gray), which is the parallel plane to the wall. Note
that $\theta$ runs from 0 (parallel to the $x$-$y$ plane) to $\pi/2$ (perpendicular to $x$-$y$
plane) given the molecule's ring symmetry and thus its otherwise degenerate orientations, such
as 0 and $\pi$. We sample the distribution of $\theta$ in each umbrella window and hence its
Boltzmann inversion, the angular potential $W(\theta,z)$. Note, that the distribution must be
normalized by the $2$sin$(\theta)$ scaling to calculate $W(\theta,z)$.

The additional residue breaks the ring symmetry which we assumed for the benzene ring.
Therefore, we formally replace our angular coordinate $\theta$ by a new angle $\phi$. It defines
the angle between the respective residue \textbf{\textsf{R}} and the $z$-axis as illustrated in
the upper sketch of Fig.~\ref{fig4}. Note that the unique mapping onto $\phi$ ranges from $0$
to $\pi$, which is the necessary descriptor range to distinguish all orientations of the ligands
other than benzene. If the angle is zero the residue points into the water, away from the
pocket. If the angle is equal to $\pi$ the residue points into the pocket, away from the bulk.
These two orientations are degenerate for the benzene ring and the definition of $\theta$, as
mentioned above. We sample the potential $U(\phi,z)$ from the orientation distributions in all
umbrella windows in the same way we obtained the potential $W(\theta,z)$. Note, that the
distributions must be normalized by the sin$(\phi)$ scaling to calculate $U(\phi,z)$.

\subsection{Unconstrained simulations for mean first passage times}

For each setup, we store a production run of 20~ns in steps of 0.2~ps in which the ligand is
constrained at the reflective boundary at $z=22$~{\AA} for the planar binding site and
$z=29$~{\AA} for the pocketed binding site. This initial trajectory served as a source for
randomly seeded initial configurations for subsequent binding event simulations. We then sampled
more than a thousand binding events starting from independent, initial frames by randomly
picking one from the previously generated production run. To ensure a selection's randomness,
upon possible re-selection, we applied an additional annealing step: within a short simulation
of 50~ps, the configuration was heated up to 350~K in a stochastic integrator scheme. After
this, the heated simulation was equilibrated for 100~ps at 298~K using the Berendsen thermostat
and the Velocity Verlet algorithm. In the final production run the ligand was released, free to
move along the $z$-direction. The runs were terminated once the ligand bound to the binding site
at $z=4$~{\AA}. The time for binding, that is, the first passage time (FPT), was then averaged
to calculate the mean first passage time (MFPT) $\mathcal{T}(z,z_f)$, the time the ligand takes
to bind from $z$ to the final bound position $z_f=4$~{\AA}. The curves for $\mathcal{T}(z,z_f)$
are discussed in the SI.

In simple cases, the MFPT curve can be theoretically calculated using a Markovian approach for
diffusion in one dimension. Given an energy landscape $V(z)$ such as a PMF and a possibly
spatially dependent friction $\xi(z)$, the MFPT can be calculated by~\cite{FirstPassageWeiss, Siegert}
\begin{equation}
\mT(z,z_f) = \beta \int_{z_f}^z \dd z' \xi(z') \mrm{e}^{\beta V(z')} \int_{z'}^{z_\mrm{max}} \dd z'' \mrm{e}^{-\beta
V(z'')} \, .
\label{eq1}
\end{equation}
where $z_f$ and $z_\mrm{max}$ denote an absorbing and reflective boundary, respectively. We
exploit and elaborate on the framework of Eq.~\eqref{eq1} in section~B, C, and D.

\section{Results}

\subsection{Ligand reorientation}

\begin{figure}[t]
\centering \includegraphics[width=8.6cm]{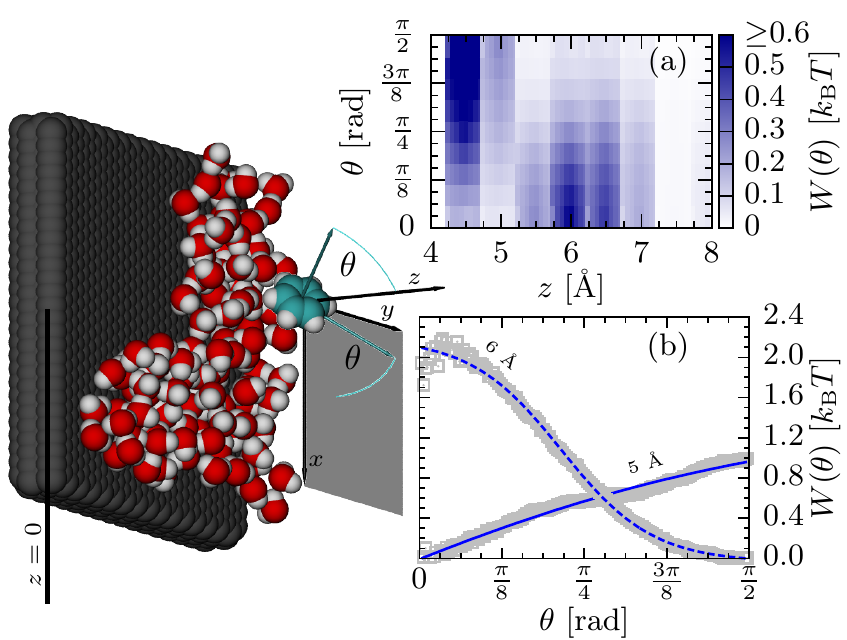} 
\caption{The simulation snapshot illustrates benzene, the planar wall and part of the water. The
binding process is constrained to one-dimensional diffusion along $z$. The ligand's orientation
is quantified by the angle $\theta$ between the atomic plane of the ring atoms and the $x$-$y$
plane (gray). It is the same angle like the one between the normal vector of benzene and the
$z$-axis. (a) The dependence of the angular potential $W(\theta,z)$ on $z$ illustrates the
pathway upon binding. Here the transition from a favorably perpendicular ($\theta=\pi/2$) to a
lateral ($\theta=0$) orientation for decreasing ligand-wall separations occurs on a narrow range
from roughly $z=5~\mrm{\AA}$ to $z=7~\mrm{\AA}$. Panel (b) exemplifies sampled data for
$W(\theta,z)$ as gray symbols including blue lined fits for $z=6~${\AA} and $z=5$~\AA.
Strikingly the perpendicular orientation is favored by over $2~k_\mrm{B}T$.}
\label{fig2}
\end{figure}

\begin{figure}[t]
\centering \includegraphics[width=8.6cm]{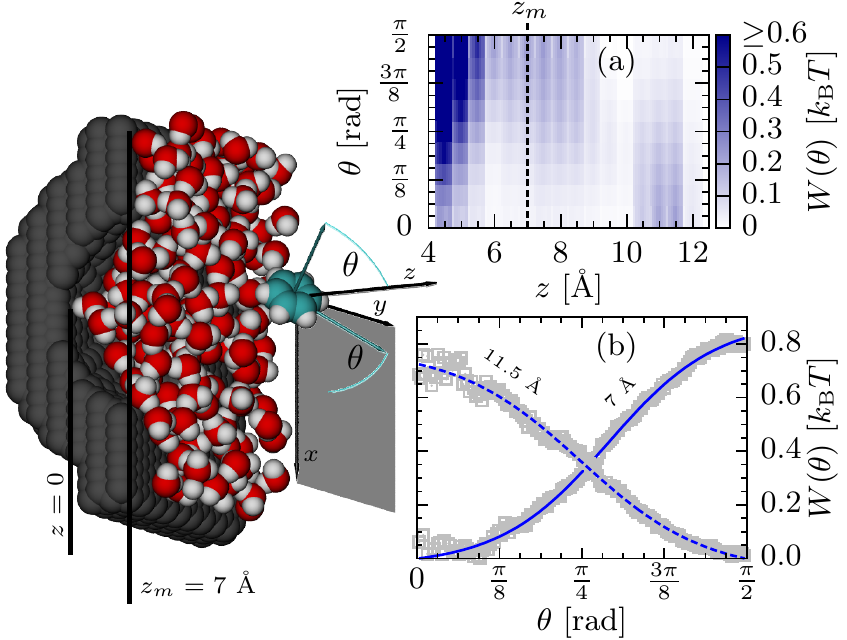}
\caption{The simulation snapshot illustrates the benzene as ligand, a section of the
hemispherical binding site and part of the water. The binding process is also constrained to
one-dimensional diffusion along $z$. The pocket bottom is defined as origin $z=0$ such that the
pocket mouth is around $z_m=7~\mrm{\AA}$. (a) The dependence of the angular potential
$W(\theta,z)$ on $z$ illustrates the pathway upon binding. For $z>10$~{\AA} a perpendicular
orientation to the wall (and towards the water interface) is energetically favored whereas for
$z<10$~{\AA} it aligns with the pocket bottom and the $x$-$y$-plane. Panel (b) exemplifies sampled
data for $W(\theta,z)$ as gray symbols including blue lined fits for $z=11.5~${\AA} and
$z=7$~\AA.}
\label{fig3}
\end{figure}

We first study the reference ligand, i.e., benzene associating to a planar hydrophobic wall.
Fig.~\ref{fig2}~(a) and (b) draw the orientation potential $W(\theta,z)$ from umbrella sampling.
Benzene undergoes a clear orientation pathway which depends on the ligand separation to the
wall. While the ligand is at $z=6$~{\AA} the perpendicular orientation for $\theta=\pi/2$ is
energetically favored by a little more than $2~k_\mrm{B}T$. The example of $W(\theta,z)$ for
$z=6$~{\AA} is shown in Fig.~\ref{fig2}~(b) where the gray symbols represent the simulation
sampled data, and the dashed blue line is a fit of a shifted hyperbolic tangent. Thus at these
ligand separations to the wall benzene partly desolvates if it orients perpendicularly to the
$x$-$y$ plane. Proceeding to smaller $z$ values aligning parallel/lateral to the wall
($\theta=0$) is favored because of steric repulsion with the wall. The example data (gray
symbols) and fit (solid blue line) of the angular potential for $z=5~${\AA} are again shown in
Fig.~\ref{fig2}~(b).

We now turn to binding to the hydrophobic pocket. 
In Fig.~\ref{fig3}~(a) and (b) we see that the benzene association to the pocketed binding site
is qualitatively similar, but the reorientation from perpendicular to lateral occurs on a much
broader range in $z$. While the ligand is $10--12$~{\AA} away from the pocket bottom, it favors the
perpendicular orientation. It is actually only slightly favored by little more than half a
$k_\mrm{B}T$ at $z=11.5$~\AA, as Fig.~\ref{fig3}~(b) exemplifies. At even closer
distances such as  $z=7$~{\AA} benzene favors the lateral orientation by more than half a
$k_\text{B}T$ as shown in panel (b). Hence, benzene favorably aligns with the pocket bottom
before it enters the pocket (compare Fig.~\ref{fig3}~(a) again).

We conclude that binding of benzene to our hydrophobic binding sites involves an energetically
favored, perpendicular orientation which is possibly disadvantageous since the final bound
configuration requires parallel alignment. For slit-like binding sites, this could be
advantageous if a ligand must enter perpendicularly before binding. We suppose that the
orientation pathway is steered by water and thus solvation free energy of the ligand because
the molecule can partly desolvate while orienting out of the water interface. In the case of a
hemispherically molded binding site, the smeared water interface allows even earlier
desolvation and reorientation upon binding which, on top, energetically weakens the
perpendicular orientation.

\begin{figure*}[t]
\centering \includegraphics{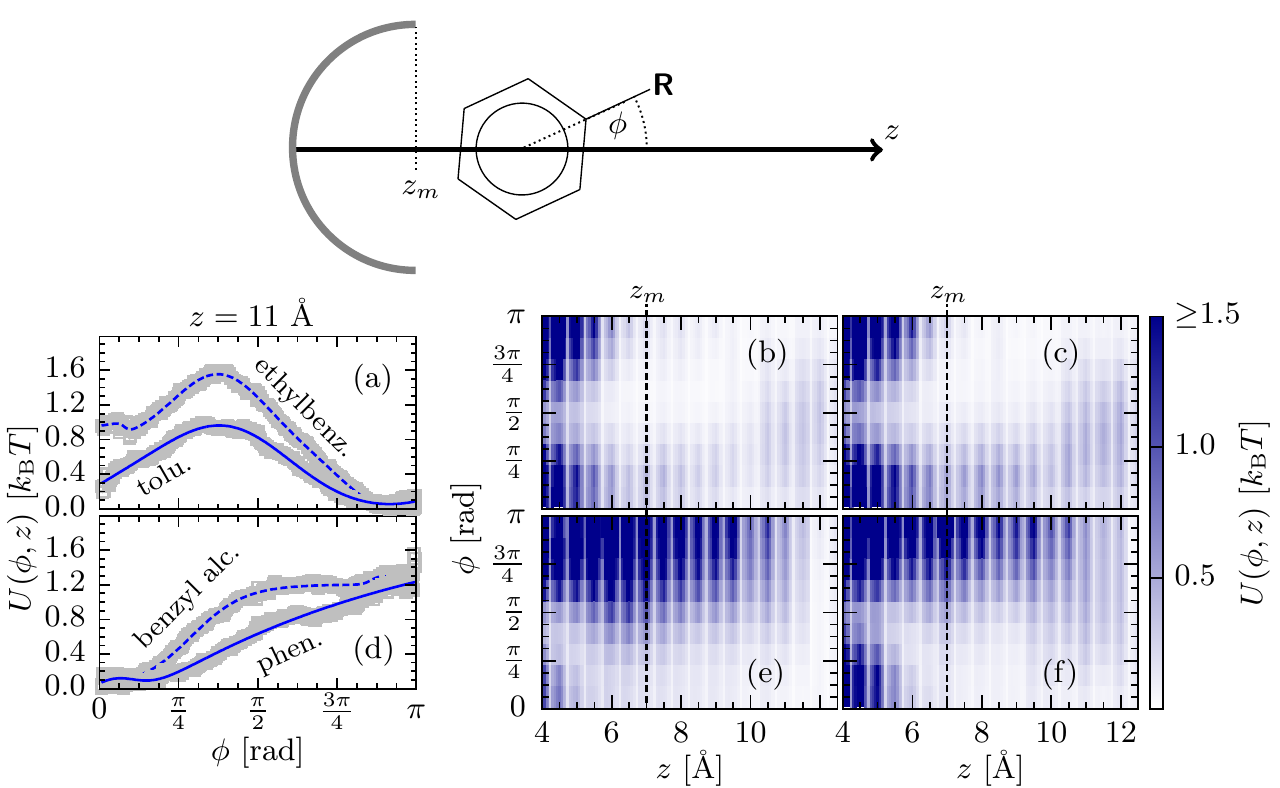}
\caption[Orientation potential for ethylbenzene, toluene, phenol and benzyl alcohol]{The upper
sketch schematically represents the pocket and ligand connected by the $z$-axis. Also the
pocket mouth $z_m=7~\mathrm{\AA}$. The angle $\phi$ is taken to be the angle between the
respective ligand's residue \textsf{R} and the $z$-axis. Plot (a) shows two examples for the
angular potentials $U(\phi,z=11~\mrm{\AA})$ from umbrella sampling of ethylbenzene (dashed) and
toluene (solid). (The gray symbols plot the sampled data and the blue lines represent smooth
interpolation functions.) The color map plots show the angular potentials from all umbrella
windows for (b) toluene and (c) ethylbenzene. Panel (d) also shows the examples of
$U(\phi,z=11~\mrm{\AA})$ for benzyl alcohol (dashed) and phenol (solid). For (e) phenol and (f)
benzyl alcohol we also show the full angular potentials $U(\phi,z)$ as color map plots.}
\label{fig4}
\end{figure*}

Additional observations on the reorientation pathways cover the association of our remaining
aromatic compounds benzyl alcohol, phenol, toluene and ethylbenzene to the pocketed binding
site only. All of these ligands comprise an aromatic ring onto which an additional residue is
attached such as the hydroxyl group to phenol.

In Fig.~\ref{fig4}~(a) two examples are shown for $U(\phi,z)$ where ethylbenzene and toluene
were constrained at $z=11$~\AA. Both exhibit a barrier around $\phi=3\pi/8$, whereas the
barrier is smaller in the case of toluene. This behavior seems to be very significant for these
two aromatic compounds which comprise an alkyl residue. Hence, if these ligands are at
intermediate positions they partly solvate either the aromatic ring (minimum at $\phi=\pi$) or
the alkyl group (minimum at $\phi=0$). Both of these orientations yield an energetic gain over
an unfavored tilted orientation around $\phi=3\pi/8$. Nevertheless, $\phi=\pi$ is globally
favored because the aromatic ring yields higher energetic contributions from the electrostatic
energy from its partial charges. Fig.~\ref{fig4}~(b) and (c) show that the bimodal
orientations of toluene and ethylbenzene range from $z=12~${\AA} to $9$~\AA. Overall the
orientation pathway funnels along angles that are larger than $\pi/2$ such that the solvation
of the aromatic ring stays favored, however, the ligand samples all orientations. Finally, in
the bound state around $z=4~\mrm{\AA}$, ethylbenzene and toluene are sterically hindered to
take orientations other than $\phi\approx\pi/2$.

The orientation pathway is again very different for phenol and benzyl alcohol. In
Fig.~\ref{fig4}~(d) the potentials $U(\phi,z=11~\mrm{\AA})$ for benzyl alcohol and phenol
exhibit a favored orientation for $\phi=0$. Both of these ligands have polar residues
which contain a hydroxyl group. These impose two potential hydrogen bonding sites: the oxygen
and the associated hydrogen atom. As a consequence, the favorable solvation of the hydroxyl
group yields a strong orientation to $\phi=0$. This angle is even more favored at closer
distances $z$, such that the barrier around $\phi=\pi$ increases to several $k_\text{B}T$. In
Fig.~\ref{fig4}~(e) and (f) for phenol and benzyl alcohol, respectively, one can see that the
energy to take an almost perpendicular orientation exceeds the $1.5~k_\mrm{B}T$ plotted scale.
Overall these plots make evident that benzyl alcohol and phenol favorably solvate their hydroxyl
groups while they approach the pocket. Finally both ligands orient to $\phi=\pi/2$ in the bound
state ($z=4~\mrm{\AA}$). In summary, the orientation of all ligands suggests that their pathways
are driven by solvation free energy such that the parts which have the highest energetic costs
upon hydration are primarily desolvated.

\begin{figure}[t]
\centering \includegraphics{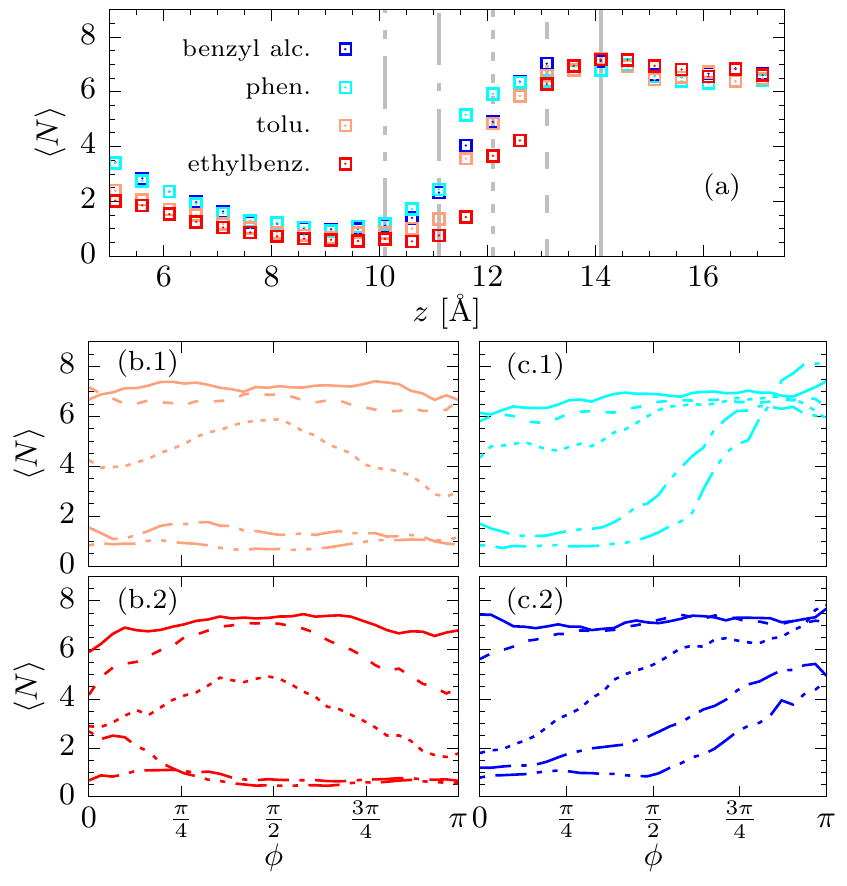}
\caption{Panel (a) shows the dewetting transition of the pocket plotting the
average pocket water occupancy against ligand distance to the pocket for benzyl alcohol, phenol,
toluene, and ethylbenzene. The gray vertical lines serve as the legend to the reference
positions that are used as line styles in the remaining figure panels. Panel (b.1) and (b.2)
plot $\langle N\rangle$ against ligand orientation $\phi$ for toluene and ethylbenzene,
respectively, at the indicated positions in panel (a). Panel (c.1) and (c.2) plot the same
$\langle N\rangle$ dependence for phenol and benzyl alcohol, respectively. Note, that while a
given ligand's hydroxyl group orients towards the pocket, the otherwise dewetted binding site
can be considerably hydrated.}
\label{fig5}
\end{figure}

In turn, we observe how ligand position and orientation influence the pocket
dewetting. Fig.~\ref{fig5} plots the average pocket water occupancy $\langle N \rangle$ against
the two spatial descriptors position $z$ and angle $\phi$ for ethylbenzene, toluene, phenol, and
benzyl alcohol. The transition from wet to dry occurs while a given ligand is at $z=10$~{\AA} to
$z=15$~{\AA} as shown in Fig.~\ref{fig5}~(a). An average of six to seven pocket water molecules
transition to less than an average of one while the ligand is still in front of the pocket at
$z=10$~{\AA}. If the ligand enters the pocket this minimum pocket water occupancy increases again
up to three to four water molecules which fit into the pocket if the ligand is bound around
$z=5$~{\AA}. Additionally, comparing the dewetting transition for instance of ethylbenzene and
phenol, we see that more elongated and hydrophobic ligands induce the dewetting farther away.
Further, the vertical gray lines indicate the reference positions for which we observe how
$\langle N\rangle$ depends on the ligand orientation in panels (b.1), (b.2), (c.1), and (c.2)
for toluene, ethylbenzene, phenol, and benzyl alcohol, respectively. In panels (b.1) and (b.2),
toluene and ethylbenzene pronouncedly induce dewetting if either elongated edge, the aromatic or
residue group, reach towards the pocket and, thus, increase the hydrophobic confinement at
$\phi=\pi$ or $\phi=0$. In contrast, phenol and benzyl alcohol enhance pocket wetting if their
polar residue is oriented towards the pocket, which essentially pushes the associated solvation
layers into the pocket. Hence the hydroxyl groups can considerably hydrate an
otherwise dewetted pocket if they are oriented toward it.

\subsection{Dissipative forces and kinetic barriers}

Previously we discussed that the pocket water density fluctuations yield additional dissipative
forces that slow the binding~\cite{Setny:PNAS, Weiss:JPCB, Weiss:JCTC}. Firstly, pocket water
occupancy fluctuations and ligand friction were shown to couple~\cite{Setny:PNAS}. The long time
transients of the water fluctuations lead to long time transients in the ligand's force
correlations, and for small ligand separations to the pocket, the water fluctuations increase
due to the increased confinement. Secondly, we derived that the hydration fluctuation time scale
and the ligand friction directly couple by a proportional relation~\cite{Weiss:JPCB}. In this
context, we found that a bimodal nature of hydration fluctuations is sufficient to enhance the
ligand friction prior to association to the pocket. And finally, we demonstrated that results
from ligand constraining simulations must be corrected by the time transients which occur in
unconstrained simulations. Only then one can capture the non-Markovian properties, i.e.
long-time correlations, for accurate kinetic predictions. We obtain the dissipative forces and
time transients (memory) in a friction profile calculated via~\cite{Hinczewski&Netz}
\begin{equation}
\beta\xi^\mrm{M}(z) = \frac{\partial \mT(z)}{\partial z}  \frac{\mrm{e}^{-\beta V(z)}}{\int_z^{z_\mrm{max}} \dd z' \mrm{e}^{-\beta
V(z')}}
\label{eq2}
\end{equation}
where the PMF $V(z)$ and the MFPT curve $\mT(z,z_f)$ are employed. Hence, we can combine the
results from ligand constraining simulations, i.e., the PMF, and unconstrained simulations,
i.e., the MFPT. We denote this profile by $\xi^\mrm{M}(z)$ accounting for the Markovian
assumption of Eq.~\eqref{eq2}. Still, we know from our previous work that this profile
non-trivially incorporates the non-Markovian memory effects by our MFPT input. Rigorously
speaking, we shall not consider $\xi^\mrm{M}(z)$ as friction profile~\cite{Setny:PNAS,
Weiss:JPCB, Weiss:JCTC}. We refer to $\xi^\mrm{M}(z)$ as the kinetic profile which can be
incorporated in a rescaled free energy landscape capturing all kinetic effects, whereas we
follow the lines of Hinczewski~\ea~\cite{Hinczewski&Netz}. 

\begin{figure}[t]
\centering{\includegraphics[width=8.6cm]{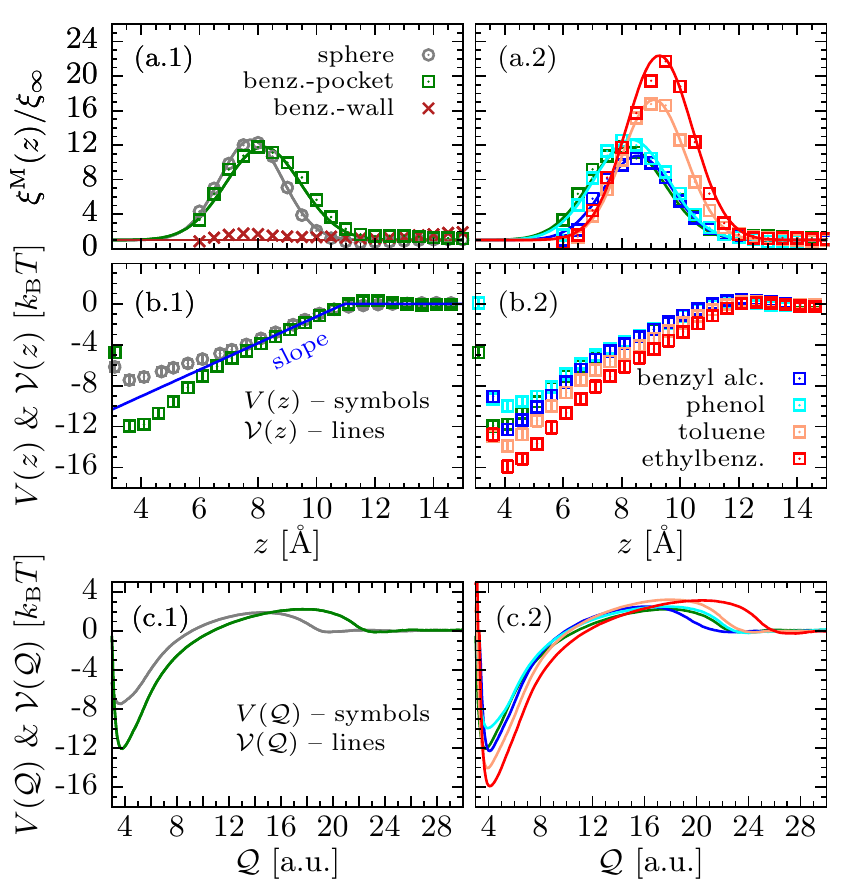}} 
\caption{(a.1) The kinetic profiles from Eq.~\eqref{eq2} exhibit peaks if benzene (green
squares) and the spherical ligand (gray circles) bind to the pocketed site. If benzene binds to
the wall (red crosses) the friction can be well assumed constant. The respectively colored lines
represent Gaussian function fits from Eq.~\eqref{eq5}. (a.2) Comparably peaking profiles can be
observed for our remaining aromatic compounds, whereas those of ethylbenzene and toluene are
even more enhanced. Panels (b.1) and (b.2) show the original PMF $V(z)$ as colored symbols. Note
that the original PMFs of benzene and the spherical ligand binding to the pocket share the
similar attracting slope (blue line) which sets in around $z=11$~\AA. Panels (c.1) and (c.2)
show the rescaled energy landscapes $\mathcal{V}(\mathcal{Q})$ as lines which exhibit the
additional kinetic barrier along the rescaled coordinate.}
\label{fig6}
\end{figure}

Fig.~\ref{fig6}~(a.1) and (a.2) show $\xi^\mrm{M}(z)/\xi_{_\infty}$ whereas we normalize by the
respective bulk friction constant to compare our various ligands. The values of the bulk
friction values are analyzed and discussed in the SI. For example, the kinetic profile of
benzene binding to the wall can be well assumed to be constant. Moreover, if benzene binds to
the pocket, the dominant feature of the steady-state friction $\xi^\mrm{M}(z)$ is a Gaussian
peaking structure, which we will model and fit by
\begin{equation}
\xi(z) = \xi_{\infty} + \Delta \xi \mathrm{e}^{-(z-z_p)^2/\sigma^2} \,\,\, .
\label{eq5}
\end{equation}
Thus, the peak height $\Delta\xi$, position $z_p$ and width $\sigma$ define a peak
that adds to the bulk friction constant $\xi_\infty$. This peak roots from pocket hydration
fluctuations as we also previously discussed for binding of a spherical ligand to the same
pocket~\cite{Weiss:JCTC} which we replot here as gray circles. The key to the additional
dissipative forces are the bimodal wet-dry hydration fluctuations which couple to the ligand.
In comparison to the data for the spherical ligand, the dissipative forces for benzene peak
wider and shift slightly further into the bulk by roughly half an \AA~ which makes their tail
reach to $z\sim11$~\AA. This well coincides with the position where the perpendicular
orientation is favored (see Fig.~\ref{fig3}~(a)). Hence, benzene is exposed to the increasing
friction at larger $z$-values because it reaches with its extended side towards the fluctuating
interface.

The kinetic profiles for the remaining aromatic compounds ethylbenzene, toluene, phenol and
benzyl alcohol are shown in Fig.~\ref{fig6}~(a.2) where they are compared to the replotted
profile of benzene. The compounds phenol and benzyl alcohol are extended by a hydroxyl group
and a methanol group, respectively, thus offering polar patches. Their size is elongated
compared to the benzene ring; however, their kinetic profiles match the one of benzene, i.e.,
their kinetic barriers well coincide. Since the orientation pathways of these two ligands
dominantly expose the aromatic ring to the pocket, the hydration fluctuations yield a similar
kinetic profile.

Ethylbenzene and toluene are purely hydrophobic compounds made up of a conjugated carbon ring
that is extended by an ethyl and a methyl group, respectively. Their kinetic profiles are more
enhanced and reach further into the bulk. For these two ligands, the peak is even higher and
hints that it contains contributions other than the bimodal pocket hydration. We suggest that
the additionally bimodally fluctuating orientation adds to the peak of the kinetic profile. In
essence, binding of these two ligands involves two degrees of freedom, which bimodally fluctuate
and which thus can both add to additional dissipative forces in our one-dimensional description.
More importantly the peak positions for ethylbenzene and toluene shift farther away from the
pocket. In comparison to benzene these ligands are even longer and are subject to the hydration
fluctuations farther outside the pocket.

To judge the impact of the kinetic profiles we rescale it into an effective free energy
landscape. We choose a new reaction coordinate $\mQ=\mQ(z)$, as suggested by
Hinczewski~\ea~\cite{Hinczewski&Netz}, such that in the new coordinates the friction is scaled
to the constant value $1~k_\mrm{B}T~\mrm{ns}~\mrm{nm}^{-2}$. Then the rescaled coordinate is
determined by $\mQ' = \dd \mQ / \dd z = \sqrt{\xi^\mrm{M}(z)/\xi_{_\infty}}$ and the PMF must be
consistently rescaled such that
\begin{equation}
\begin{split}
\mV (\mQ(z)) &= V(z) +  (\beta)^{-1} \mrm{ln}(\mQ'(z)) \\
						 &= V(z) + (2\beta)^{-1} \mrm{ln}(\xi^\mrm{M}(z)/\xi_{_\infty}) 
\end{split}
\label{eq6}
\end{equation}
In panels (c.1) and (c.2) we plot the rescaled potentials against the new coordinate $\mQ$. The
rescaled energy landscapes exhibit additional \textit{kinetic barriers} which naturally origin
from the kinetic profiles. The rescaled coordinate is calculated as the integral over
$\sqrt{\xi^\mrm{M}(z)/\xi_{_\infty}}$ such that the integration of the Gaussian shaped peak
stretches the reaction coordinate. In comparison to the case for the spherical ligand (gray
line), the peak in the kinetic barriers for the aromatic compounds (colored lines) shift farther
away from the pocket, namely to increasing values of $\mQ$. In general, the farther the barrier
shifts down the attracting slope, the smaller is its impact because the anyhow attracting slope
diminishes the repulsive slope on the r.h.s. of the barrier. In other words, part of the
repulsive slope of the kinetic barrier reaches across the onset of attraction which makes its
effect more significant for a slowed association. This result is consistent with the MFPT data
which we present in the SI, where benzene and other aromatic compounds bind slightly slower than
the spherical ligand. The binding times of ethylbenzene and toluene are even slower than those
of benzene. The binding speeds of phenol and benzyl alcohol, however, are similar to the binding
times of benzene.

In sum, the size and nature of a ligand can shift and tune the dissipative forces and the
resulting kinetic barriers in the $\xi^\mrm{M}(z)$ profiles. So far we only discussed this
qualitatively with a scientist's intuition for the shapes of energy landscapes. In the
following, we approach our arguments in a quantitative picture by which we explore the full
range of the possible impact of the kinetic barrier.

\subsection{Impact of Steady-State Friction}

\begin{figure}[t]
\includegraphics[width=8.6cm]{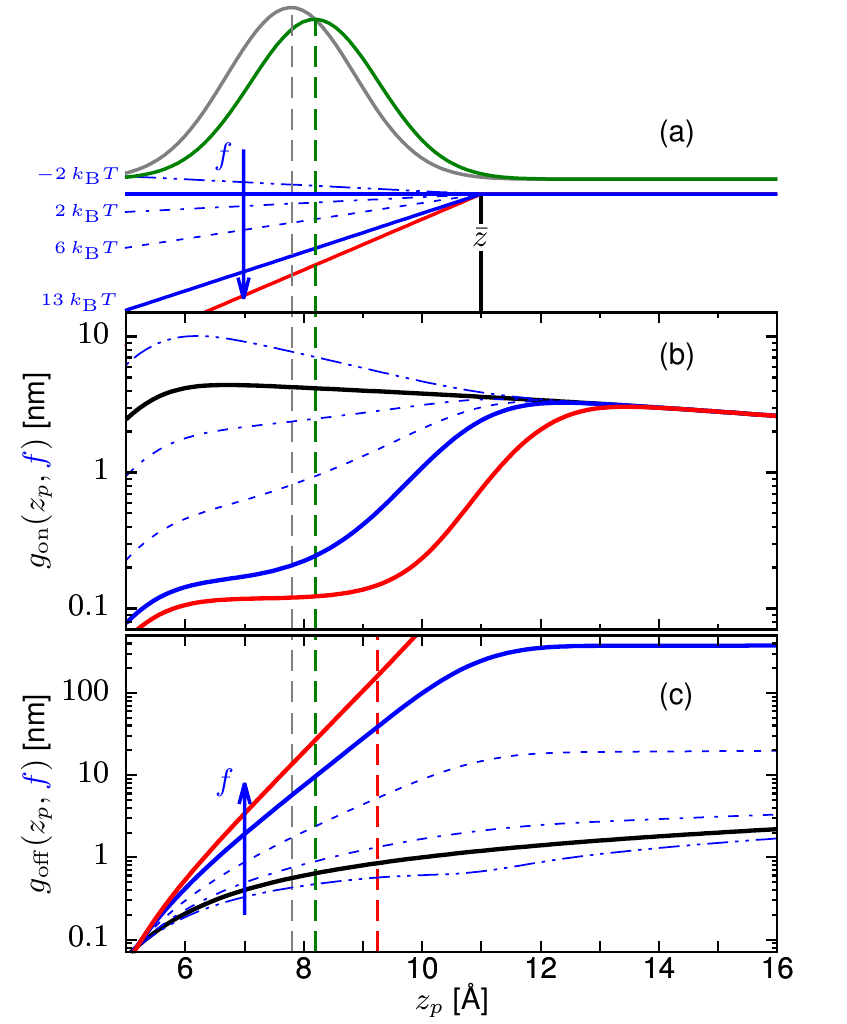}
\caption{(a) The dominant features of the ligand binding process are the potential
$V(z)=f(z-\zbar)$ (blue), that strongly attracts the ligand given $z\leq\zbar$, and a
Gaussian friction peak (gray and green) modeled by Eq.~\eqref{eq5}. Various other
potential slopes are sketched as thin blue lines. (b) The factor $g_\mathrm{on}(z_p,f)$ in
Eq.~\eqref{eq8} strongly depends on the friction peak position. While the negative shift
$z_p-\zbar$ decreases, $g_\mathrm{on}(z_p,f)$ and thus the impact of the friction peak decreases.
Moreover turning to (even slightly) repulsive slopes $f$ drastically increases the
impact of the kinetic barrier (double dotted dashed). (c) The factor
$g_\mathrm{off}(z_p,f)$ for the unbinding process exponentially increases with friction peak
position $z_p$ and the repulsive slope $f$. In the example of ethylbenzene (red curve), the
potential slope is the steepest, and the friction peak position lies farthest outside the pocket
which is why the scaling factor is on the order of $\mathcal{O}(10^2)$.}
\label{fig7}
\end{figure}

We already formalized the kinetic profile $\xi^\mrm{M}(z)$ by its dominant feature the fitted
Gaussian peak from Eq.~\eqref{eq5} (see also Fig.~\ref{fig6}). The common and dominant feature
of the PMF is its significantly attracting slope with roughly $f=13~k_\mrm{B}T~\mrm{nm}^{-1}$ (see
blue line in Fig.~\ref{fig6}~(b.1)). Thus in the following minimalistic model we use
$V(z)=f(z-\bar{z})$ for $z\leq\bar{z}$ and $V(z)=0$ otherwise, such that $\zbar$ denotes the
inset position of a constant attraction with strength $f$. For illustration the simplified
potential and friction are plotted together in the upper sketch of Fig.~\ref{fig7}. In particular,
the contribution of the friction peak, i.e., the second summand on the r.h.s. of
Eq.~\eqref{eq5}, to the binding time in Eq.~\eqref{eq1} is given by
\begin{equation}
\Delta\mT = \beta \Delta\xi \int_{z_f}^z \dd z'\mexp^{\beta V(z')-\frac{(z'-z_p)^2}{\sigma^2}} \int_{z'}^{z_\mathrm{max}} \dd z'' \mexp^{-\beta V(z'')}
\label{eq7}
\end{equation}
whereas the stepwise definition of $V(z)$ has yet to be evaluated.

Fixing $\zbar=11$~\AA, $z_f=4$~\AA,and $z_\mrm{max}=29$~\AA, leaves the bracket in the integral
in Eq.~\ref{eq7} dependent on the friction peak position $z_p$, width $\sigma$ and the force
constant $f$. We lay out the detailed integration steps in the Appendix~A. The result simplifies
to
\begin{equation}
\Delta\mT = \frac{\sqrt{\pi}\sigma}{2} \beta\Delta\xi \cdot
g_\mathrm{on}(z_p, f)
\label{eq8}
\end{equation}
the product of friction peak height, width and a scaling factor
$g_\mathrm{on}(z_p,f)$. For the moment we can neglect the dependence of
$g_\mathrm{on}(z_p,f)$ on $\sigma$ because the direct proportionality of
$\Delta T\propto\sigma$ is the dominating peak width dependence for our values of $\sigma$. The
factor $g_\mathrm{on}(z_p,f)$ quantifies the impact of the friction peak on the binding time which is why we will also refer to it as the scaling factor.

If we choose the slope of $f=13~k_\mrm{B}T~\text{nm}^{-1}$ from Fig.~\ref{fig6}~(b.1), the
scaling factor, shown as blue solid line in Fig.~\ref{fig7}~(b), steeply increases with the peak
position of the kinetic profile. The broken blue line types indicate how $g_\mathrm{on}(z_p,f)$ increases
with decreasing potential slope $f$ whereas the thick black line is the case for $f=0$. If the
force constant even becomes repulsive the scaling factor increases drastically because the
repulsive potential slope and the repulsive kinetic barrier add up. We find that
the mean binding time is certainly affected by and thus proportional to the friction peak height
$\Delta\xi$, although, the impact can drastically decrease if the peak shifts downward the
attracting slope. For repulsive slopes the scaling can drastically
dominate such that for our case the unbinding is dominantly affected.

The result of the scaling factor $g_\mathrm{off}$ for the unbinding process is
determined by interchanging the integration boundaries $z_{max}$ and $z_f$ in Eq.~\eqref{eq7}.
We exemplify the scaling factor for the unbinding in Fig.~\ref{fig7}~(c). Generally, the scaling
factors exponentially scale with force constant $f$. Especially for our linearly attractive
potential, they scale with $g_\mathrm{on}\propto\mrm{exp}(\beta f
(\bar{z}-z_f))$ for the binding process and
$g_\mathrm{off}\propto\mrm{exp}(\beta f(z_p-\bar{z}))$ for the unbinding
process. Thus for unbinding it takes values much larger than one, i.e., $\mathcal{O}(10^1)$, where
for the binding it takes values two orders of magnitude smaller, i.e., $\mathcal{O}(10^{-1})$.
In the special case of ethylbenzene, the slope $f$ is the steepest and the friction peak is
farthest away from the pocket. Hence, the scaling factor is on the order of
$\mathcal{O}(10^2)$ for ethylbenzene. See also red curve in Fig.~\ref{fig7}~(c).

\subsection{Unbinding}

\begin{figure}[b]
\includegraphics[width=8.6cm]{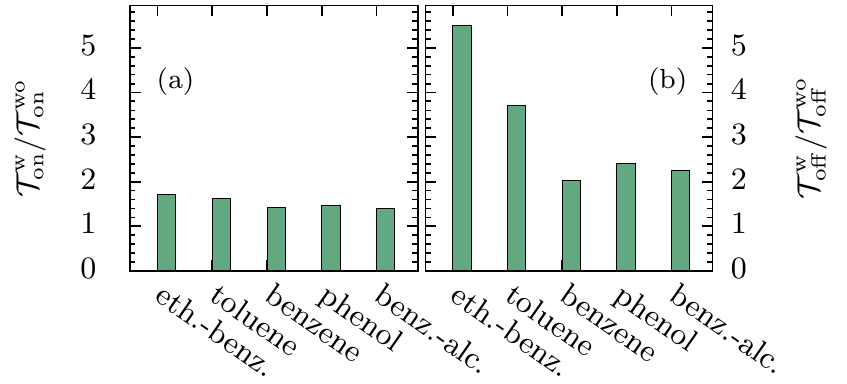}
\caption{(a) The ratio of the average binding time with $\mT_\mrm{on}^\mrm{w}$ and without
$\mT_\mrm{on}^\mrm{wo}$ kinetic barrier exhibits a constant impact during binding. (b) In
contrast, the unbinding times dominantly increase by a factor of five in the ratio of the
average unbinding times with $\mT_\mrm{off}^\mrm{w}$ and without $\mT_\mrm{off}^\mrm{wo}$
kinetic barrier.}
\label{fig8}
\end{figure}

In this section we want to compare the impact on average binding times
\begin{equation}
\mT_\mrm{on} = \frac{\int_{z_\mrm{b}}^{z_\mrm{max}} \dd z' \mT(z',z_\mrm{b}) }{(z_\mrm{max}-z_\mrm{b})} 
\label{eq11}
\end{equation}
where we choose the boundary to be bound as $z_\mrm{b}=10$~\AA, and the average unbinding times 
\begin{equation}
\mT_\mrm{off} = \frac{\int_{z_\mrm{ub}}^{z_\mrm{f}} \dd z' \mT(z',z_\mrm{ub}) }{(z_\mrm{max}-z_\mrm{ub})} 
\label{eq12}
\end{equation}
where we choose the boundary to be unbound as $z_\mrm{ub}=12$~\AA. The boundary for binding is
read from the kinetic barrier peak position. The boundary for the unbinding $z_\mrm{ub}$ is
chosen such that the unbinding process can be considered complete by overcoming the kinetic
barrier. The respective MFPT curve $\mT(z,z_\mrm{b/ub})$ is calculated from Eq.~\eqref{eq1}
where the upper integration boundary is $z_\mrm{max}$ for the binding case and $z_f$ for the
unbinding case. For each ligand we choose two scenarios -- one neglecting the kinetic barrier,
thus $\xi(z) = \xi_\infty$ and one incorporating the kinetic barrier. 

The histogram of Fig.~\ref{fig8}~(a) plots the ratio
$\mT^\mrm{w}_\mrm{on}/\mT^\mrm{wo}_\mrm{on}$ of the binding time with and without kinetic
barrier. The barrier slows the binding time by a factor smaller than two for all ligands. In
contrast, the average unbinding time is dominantly affected. We estimate that the kinetic
barrier adds an extra $221~\mu$s to the unbinding time of ethylbenzene and less than $1~\mu$s to
that of phenol. Moreover, the ratio of the average unbinding times \textit{with} and
\textit{without} kinetic barrier in Fig.~\ref{fig8}~(b) yields a factor of more than five for
ethylbenzene and is generally non-constant for our various ligands. Note, however, that the
proper kinetic barrier for the unbinding process can differ from the $\xi^\text{M}(z)$-profiles
which we originally extracted from the binding process. Hence we neglect possible hysteresis
effects. Nevertheless, our procedure is most sufficient and efficient for the conclusive
interpretation of our estimates of the unbinding times. Thus, we assume that the conclusions and
implications remain the same since, in particular, the energy landscape or binding affinity
mainly steers the scaling of the kinetic barrier, while height modulations of the barriers
linearly influence the average unbinding time (compare Eq.~\eqref{eq8} again).

On the one hand, the resulting estimates for the unbinding time generally confirm that more
(extended) hydrophobic ligands reside longer inside the pocket. On the other hand, if we neglect
the additional kinetic barrier in front of the pocket, the unbinding estimates can be wrong by
several hundred microseconds. In particular, ethylbenzene would actually stay for
$270~\mu\mrm{s}$ if estimated by Eq.~\eqref{eq1} incorporating the kinetic barrier, while it
would only stay for $49~\mu\mrm{s}$ if the kinetic barrier is ignored. In summary, we find a
constant impact on the binding times, while the impact on unbinding times predominantly
changes.

\section{Conclusion}

The ubiquitous motifs of hydrophobic and hydrophilic groups in active compounds are undoubtedly
recognized in biomedical applications and the optimization of the overall efficacy of \textit{in
vitro} and \textit{in vivo} systems. One increasingly appreciated aspect of the optimization is
the kinetics of association and dissociation. In this regard, solvent-mediated interactions,
offer novel possibilities for control of synthetic cavitands and drug discovery.

In summary, we investigated how binding site hydration influences the binding and unbinding
kinetics of aspherical, i.e., aromatic, ligands. Therefore, we compared binding of benzene to
two different binding sites: our hydrophobic pocket and a planar wall. We found that the benzene
ring intermediately oriented such that it could maximally desolvate. The aromatic ring took a
perpendicular orientation to the binding site if it was just about to enter the pocket. In
contrast, if it was bound it favored a laterally aligned, i.e., flat, orientation to the binding
site. The general rule, however, was apparent from adding the observations of other aromatic
compounds, i.e., ethylbenzene, toluene, phenol, and benzyl alcohol. The ligands would undergo a
reorientation process that seemed to be driven by solvation free energy. Hydrophobic groups
would primarily desolvate by orienting towards the water interface. In the cases of ethylbenzene
and toluene, the energetically favored orientation even became bimodal such that two distinct
orientations were locally stable. In all cases, the aromatic ligands underwent a reorientation
process in which an intermediate orientation was orthogonal to the orientation of the final
bound state. We found that concerning this orthogonal reorientation it could be advantageous to
bind to a dewetted pocket in comparison of binding to a wall because the energetic penalties for
reorientation were more moderate in the pocket case. Additionally, the whole reorientation
process was stretched over a broader spatial range also because the pocket was strongly
dewetted.

These findings complement on previous, more coarse-grained, studies about the solvent-mediated
depletion potentials and entropically driven torque from density functional theory
(DFT)~\cite{Roth, König}. These studies on the solvent-mediated association of ideal solutes
revealed an association pathway which exhibited an intermediately tilted orientations of the
extended, aspherical solutes -- neither orthogonal nor laterally aligned. In particular, a
ligand would thus approach a binding site with a relatively tilted orientation and then lay down
into the pocket~\cite{König}. In contrast to these DFT studies, we found that for our systems a
hydrophobically driven torque orients the ligand perpendicular to the binding site, which can be
considered 'disadvantageous' if the bound state requires the ligand to be parallely aligned to
the binding site. The perpendicular orientation could be considered 'advantageous', if the
ligand has to enter a narrow, slit-like, elliptical pocket. Moreover, a solvation driven torque
should comprise entropic as well as enthalpic contributions in comparison to the aforementioned
entropic torque. In sum, the 'solvation' or 'hydrophobic torque' originating from solvation free
energy is steered by specific chemical groups of the ligands where the specific behavior of a
respective ligand can often be well anticipated such that they imply simple design principles
for steered orientation pathways.

Additionally, we studied the kinetics by asking how the kinetic coupling impacts the binding and
also the unbinding times. We compared the various ligands to our previous results of the
spherical ligand from Ref.~\cite{Weiss:JCTC}. The aromatic compounds were binding slower, when
we compared binding times that were normalized by the bulk friction coefficients, even though
some of them exhibited a stronger binding affinity than the sphere. For the discussion of these
findings, we returned to the approach of Eq.~\eqref{eq1} to extract the kinetic profiles via
Eq.~\eqref{eq2}. By definition, the kinetic profiles reproduced the correct mean binding times.
Nevertheless, we could reinterpret the peaks from dissipative forces as kinetic barriers which
scaled into new energy landscapes using a methodology depicted in Ref.~\cite{Hinczewski&Netz}.

We rationalized the effect of the kinetic barrier regarding a scaling factor which dominantly
depended on the barrier’s position relative to the anyhow attractive slope of the respective
PMF. We found that the dissipative forces can have a much higher impact on unbinding times. The
binding times were similarly enhanced by various kinetic barriers for the different ligands;
however, the effect on unbinding times scaled from a factor from less than two to five. In
particular, the residence time could be extended by hundreds of microseconds if the theoretical
estimates accounted for the kinetic barrier. This slow down was especially pronounced for the
ligands for which the orientation fluctuated bimodally in front of the pocket. In general, the
additional degree of freedom of pocket hydration adds to dissipative forces that are not
captured inside a PMF~\cite{Weiss:JPCB}. In this context, we suggest that other bimodal degrees
of freedom can add to the kinetic barrier such that the bimodally fluctuating orientation of
ethylbenzene and toluene could also increase the effective friction. Nevertheless, the major
impact of the kinetic barriers on unbinding times was steered by the slope in the energy
landscapes and how far the barriers reached outside the pocket. In that respect, extended
ligands proved to be appropriate to shift the peak away from the pocket while the extended side
of elongated compounds orients towards the pocket and induced solvation fluctuations while the
ligand is even farther outside.

As a final notion, we highlight again that one significant model restriction is the
one-dimensional treatment along the $z$-coordinate. In particular, Tiwary et
al.~\cite{Tiwary:PNAS} critically assessed the one-dimensional restraint in a similar MD setup,
where they found that the ligand least likely enters via a pathway that would include enhanced
water fluctuations. This stands in line with our interpretation of the enhanced friction as
kinetic barriers. Consequently, a ligand might particularly avoid a route comprising kinetic
barriers which in turn guide the possible pathways in a given system. Our model enables a
purified and idealized investigation regarding mechanisms which are certainly not exclusively
restricted to this ligand-pocket setup. Similar results from other model solutes similarly infer
the far-reaching consequences~\cite{MorroneLiBerne, LiMorroneBerne}. Hence, we focus on this
model system to investigate water fluctuation driven effects which system-dependently influence
association kinetics. We leave the assessment to which extent water fluctuations play a role in
a given system to future studies which especially deal with realistic association processes.
Further, the presence and relevance of drying transitions in free energy pathways have also been
emphasized in the context of folding and function of proteins. The kinetics in protein folding
has previously been explored by Hinczewski et al.~\cite{Hinczewski&Netz} when they introduced
the aforementioned rescaling procedure. One of their main conclusions was that the importance of
novel features in the rescaled energy landscape especially increased due to explicit water
effects which introduced new kinetic mechanisms. Our study is oriented along these lines of a
fundamental understanding of how solvation impacts kinetic mechanisms while the model nature of
our setup yields the insights for fundamental relationships and controllability.

\textbf{Supporting information}

Details on the bulk friction constants $\xi_\infty$, MFPT, and PMF.

\begin{acknowledgements}
The authors thank the Deutsche Forschungsgemeinschaft (DFG) for financial support
for this project. P.S. is supported by EMBO IG 3051/2015.
\end{acknowledgements}

\appendix
\section{Calculation of scaling factor $\bm{g}_\text{on}$}

Piecewise evaluation of the inner integral $\mathcal{I}(z)$ comprises a trivial case, i.e. an
integral over unity, while $z>\zbar$ and an integral over the Boltzmann factor
$\mathrm{e}^{-\beta V(z'')}$ while $z\leq\zbar$. The result thus remains piecewise defined,
such that
\begin{equation}
\mathcal{I}(z) = 
\begin{cases}
\frac{1}{\beta f} \left [ \mexp^{-\beta f(z'-\bar{z})} - 1 \right ] + (z_\mrm{max} -
\bar{z} ) & \text{for}~z' < \bar{z} \\
 z_\mrm{max} - z' & \text{for}~z' > \bar{z}
\end{cases}
\label{eqS4}
\end{equation}
(Note that the piecewise definition $\mathcal{I}(z)$ would be lost if $V(z)$ would have been
discretized by a Heaviside Step function.) In the next step the outer integral is
\begin{widetext}
\begin{subequations}
\begin{eqnarray}
\Delta \mT \propto && \int_{z_f}^z \dd z'\mexp^{-(z'-z_p)^2/\sigma^2} \mexp^{\beta V(z')} \times \mathcal{I}(z') \label{eqS51} \\
= && \int_{z_f}^{\bar{z}} \dd z'\mexp^{-(z'-z_p)^2/\sigma^2} \left [ \frac{1-\mexp^{\beta f(z'-\bar{z})}}{\beta f} +
(z_\mrm{max} - \bar{z}) \mexp^{\beta f(z' - \bar{z})}\right ] \label{eqS52} \\
&+& \int_{\bar{z}}^z \dd z'\mexp^{-(z'-z_p)^2/\sigma^2} \times (z_\mrm{max} - z') \label{eqS53}
\end{eqnarray}
\end{subequations}
where we used the piecewise definitions of $V(z)$ and $\mathcal{I}(z)$ to split the integral
from $z_f$ to $z$ into an integral from $z_f$ to $\zbar$ and another one from $\zbar$ to $z$.
Additionally the inverse Boltzmann factor $\mathrm{e}^{\beta V(z')}$ in Eq.~\eqref{eqS51} is
pulled into the square brackets in Eq.~\eqref{eqS52} ( and is one in Eq.~\eqref{eqS53}).
Completing the squares, if necessary, all integrals can be related to a Gaussian/Euler-Poisson
integral. If we neglect $\mathcal{O}(\sigma^2)$ terms the solution reads
\begin{subequations}
\begin{eqnarray}
\Delta \mT = \frac{\sqrt{\pi}\sigma}{2} \beta \Delta\xi  \Big\{ && \frac{1}{\beta f} \left ( \erf \left [ \frac{z_p - z_f}{\sigma} \right ] - \erf \left [ \frac{z_p-\bar{z}}{\sigma} \right ] \right ) \\
&+& \left ( -\frac{1}{\beta f} + (z_\mrm{max}-\bar{z}) \right ) \mexp^{\beta f (z_p-\bar{z})} \erf\left [ \frac{\beta f
\sigma}{2} + \frac{z_p-z_f}{\sigma} \right ] \\ 
&-& \left ( -\frac{1}{\beta f} + (z_\mrm{max}-\bar{z}) \right ) \mexp^{\beta f (z_p-\bar{z})} \erf\left [ \frac{\beta f
\sigma}{2} + \frac{z_p-\zbar}{\sigma} \right ] \\
&+& (z_\mrm{max} - z_p) \left ( \erf \left [ \frac{z_p-\zbar}{\sigma} \right ] - \erf \left [\frac{z_p -
z}{\sigma} \right ] \right ) \,\,\,~\Big\}
\end{eqnarray}
\label{eqS6}
\end{subequations}
\end{widetext}


\begin{thebibliography}{46}%
\makeatletter
\providecommand \@ifxundefined [1]{%
 \@ifx{#1\undefined}
}%
\providecommand \@ifnum [1]{%
 \ifnum #1\expandafter \@firstoftwo
 \else \expandafter \@secondoftwo
 \fi
}%
\providecommand \@ifx [1]{%
 \ifx #1\expandafter \@firstoftwo
 \else \expandafter \@secondoftwo
 \fi
}%
\providecommand \natexlab [1]{#1}%
\providecommand \enquote  [1]{``#1''}%
\providecommand \bibnamefont  [1]{#1}%
\providecommand \bibfnamefont [1]{#1}%
\providecommand \citenamefont [1]{#1}%
\providecommand \href@noop [0]{\@secondoftwo}%
\providecommand \href [0]{\begingroup \@sanitize@url \@href}%
\providecommand \@href[1]{\@@startlink{#1}\@@href}%
\providecommand \@@href[1]{\endgroup#1\@@endlink}%
\providecommand \@sanitize@url [0]{\catcode `\\12\catcode `\$12\catcode
  `\&12\catcode `\#12\catcode `\^12\catcode `\_12\catcode `\%12\relax}%
\providecommand \@@startlink[1]{}%
\providecommand \@@endlink[0]{}%
\providecommand \url  [0]{\begingroup\@sanitize@url \@url }%
\providecommand \@url [1]{\endgroup\@href {#1}{\urlprefix }}%
\providecommand \urlprefix  [0]{URL }%
\providecommand \Eprint [0]{\href }%
\providecommand \doibase [0]{http://dx.doi.org/}%
\providecommand \selectlanguage [0]{\@gobble}%
\providecommand \bibinfo  [0]{\@secondoftwo}%
\providecommand \bibfield  [0]{\@secondoftwo}%
\providecommand \translation [1]{[#1]}%
\providecommand \BibitemOpen [0]{}%
\providecommand \bibitemStop [0]{}%
\providecommand \bibitemNoStop [0]{.\EOS\space}%
\providecommand \EOS [0]{\spacefactor3000\relax}%
\providecommand \BibitemShut  [1]{\csname bibitem#1\endcsname}%
\let\auto@bib@innerbib\@empty
\bibitem [{\citenamefont {Kim}\ \emph {et~al.}(1997)\citenamefont {Kim},
  \citenamefont {Lew}, \citenamefont {Williams}, \citenamefont {Liu},
  \citenamefont {Zhang}, \citenamefont {Swaminathan}, \citenamefont
  {Bischofberger}, \citenamefont {Chen}, \citenamefont {Mendel}, \citenamefont
  {Tai}, \citenamefont {Laver},\ and\ \citenamefont {Stevens}}]{enzyme1}%
  \BibitemOpen
  \bibfield  {author} {\bibinfo {author} {\bibfnamefont {C.~U.}\ \bibnamefont
  {Kim}}, \bibinfo {author} {\bibfnamefont {W.}~\bibnamefont {Lew}}, \bibinfo
  {author} {\bibfnamefont {M.~A.}\ \bibnamefont {Williams}}, \bibinfo {author}
  {\bibfnamefont {H.}~\bibnamefont {Liu}}, \bibinfo {author} {\bibfnamefont
  {L.}~\bibnamefont {Zhang}}, \bibinfo {author} {\bibfnamefont
  {S.}~\bibnamefont {Swaminathan}}, \bibinfo {author} {\bibfnamefont
  {N.}~\bibnamefont {Bischofberger}}, \bibinfo {author} {\bibfnamefont {M.~S.}\
  \bibnamefont {Chen}}, \bibinfo {author} {\bibfnamefont {D.~B.}\ \bibnamefont
  {Mendel}}, \bibinfo {author} {\bibfnamefont {C.~Y.}\ \bibnamefont {Tai}},
  \bibinfo {author} {\bibfnamefont {W.~G.}\ \bibnamefont {Laver}}, \ and\
  \bibinfo {author} {\bibfnamefont {R.~C.}\ \bibnamefont {Stevens}},\
  }\href@noop {} {\bibfield  {journal} {\bibinfo  {journal} {J.~Am.~Chem.~So.}\
  }\textbf {\bibinfo {volume} {119}},\ \bibinfo {pages} {681} (\bibinfo {year}
  {1997})}\BibitemShut {NoStop}%
\bibitem [{\citenamefont {Creese}\ \emph {et~al.}(1976)\citenamefont {Creese},
  \citenamefont {Burt},\ and\ \citenamefont {Snyder}}]{dopamine}%
  \BibitemOpen
  \bibfield  {author} {\bibinfo {author} {\bibfnamefont {I.}~\bibnamefont
  {Creese}}, \bibinfo {author} {\bibfnamefont {D.~R.}\ \bibnamefont {Burt}}, \
  and\ \bibinfo {author} {\bibfnamefont {S.~H.}\ \bibnamefont {Snyder}},\
  }\href@noop {} {\bibfield  {journal} {\bibinfo  {journal} {Science}\ }\textbf
  {\bibinfo {volume} {192}},\ \bibinfo {pages} {481} (\bibinfo {year}
  {1976})}\BibitemShut {NoStop}%
\bibitem [{\citenamefont {Borngraeber}\ \emph {et~al.}(2003)\citenamefont
  {Borngraeber}, \citenamefont {Budny}, \citenamefont {Chiellini},
  \citenamefont {Cunha-Lima}, \citenamefont {Togashi}, \citenamefont {Webb},
  \citenamefont {Baxter}, \citenamefont {Scanlan},\ and\ \citenamefont
  {Fletterick}}]{hormone1}%
  \BibitemOpen
  \bibfield  {author} {\bibinfo {author} {\bibfnamefont {S.}~\bibnamefont
  {Borngraeber}}, \bibinfo {author} {\bibfnamefont {M.-J.}\ \bibnamefont
  {Budny}}, \bibinfo {author} {\bibfnamefont {G.}~\bibnamefont {Chiellini}},
  \bibinfo {author} {\bibfnamefont {S.~T.}\ \bibnamefont {Cunha-Lima}},
  \bibinfo {author} {\bibfnamefont {M.}~\bibnamefont {Togashi}}, \bibinfo
  {author} {\bibfnamefont {P.}~\bibnamefont {Webb}}, \bibinfo {author}
  {\bibfnamefont {J.~D.}\ \bibnamefont {Baxter}}, \bibinfo {author}
  {\bibfnamefont {T.~S.}\ \bibnamefont {Scanlan}}, \ and\ \bibinfo {author}
  {\bibfnamefont {R.~J.}\ \bibnamefont {Fletterick}},\ }\href@noop {}
  {\bibfield  {journal} {\bibinfo  {journal} {Proc.~Natl.~Acad.~Sci.~USA}\
  }\textbf {\bibinfo {volume} {100}},\ \bibinfo {pages} {15358} (\bibinfo
  {year} {2003})}\BibitemShut {NoStop}%
\bibitem [{\citenamefont {Teixeira-Clerc}\ \emph {et~al.}(2006)\citenamefont
  {Teixeira-Clerc}, \citenamefont {Julien}, \citenamefont {Grenard},
  \citenamefont {Van~Nhieu}, \citenamefont {Deveaux}, \citenamefont {Li},
  \citenamefont {Serriere-Lanneau}, \citenamefont {Ledent}, \citenamefont
  {Mallat},\ and\ \citenamefont {Lotersztajn}}]{cb1}%
  \BibitemOpen
  \bibfield  {author} {\bibinfo {author} {\bibfnamefont {F.}~\bibnamefont
  {Teixeira-Clerc}}, \bibinfo {author} {\bibfnamefont {B.}~\bibnamefont
  {Julien}}, \bibinfo {author} {\bibfnamefont {P.}~\bibnamefont {Grenard}},
  \bibinfo {author} {\bibfnamefont {J.~T.}\ \bibnamefont {Van~Nhieu}}, \bibinfo
  {author} {\bibfnamefont {V.}~\bibnamefont {Deveaux}}, \bibinfo {author}
  {\bibfnamefont {L.}~\bibnamefont {Li}}, \bibinfo {author} {\bibfnamefont
  {V.}~\bibnamefont {Serriere-Lanneau}}, \bibinfo {author} {\bibfnamefont
  {C.}~\bibnamefont {Ledent}}, \bibinfo {author} {\bibfnamefont
  {A.}~\bibnamefont {Mallat}}, \ and\ \bibinfo {author} {\bibfnamefont
  {S.}~\bibnamefont {Lotersztajn}},\ }\href@noop {} {\bibfield  {journal}
  {\bibinfo  {journal} {Nat. Med.}\ }\textbf {\bibinfo {volume} {12}},\
  \bibinfo {pages} {671} (\bibinfo {year} {2006})}\BibitemShut {NoStop}%
\bibitem [{\citenamefont {Hillyer}\ \emph {et~al.}(2016)\citenamefont
  {Hillyer}, \citenamefont {Gibb}, \citenamefont {Sokkalingam}, \citenamefont
  {Jordan}, \citenamefont {Ioup},\ and\ \citenamefont {Gibb}}]{Gibb:OrgLett}%
  \BibitemOpen
  \bibfield  {author} {\bibinfo {author} {\bibfnamefont {M.~B.}\ \bibnamefont
  {Hillyer}}, \bibinfo {author} {\bibfnamefont {C.~L.~D.}\ \bibnamefont
  {Gibb}}, \bibinfo {author} {\bibfnamefont {P.}~\bibnamefont {Sokkalingam}},
  \bibinfo {author} {\bibfnamefont {J.~H.}\ \bibnamefont {Jordan}}, \bibinfo
  {author} {\bibfnamefont {S.~E.}\ \bibnamefont {Ioup}}, \ and\ \bibinfo
  {author} {\bibfnamefont {B.~C.}\ \bibnamefont {Gibb}},\ }\href@noop {}
  {\bibfield  {journal} {\bibinfo  {journal} {Org.~Lett.}\ }\textbf {\bibinfo
  {volume} {18}},\ \bibinfo {pages} {4048} (\bibinfo {year}
  {2016})}\BibitemShut {NoStop}%
\bibitem [{\citenamefont {Wanjari}\ \emph {et~al.}(2016)\citenamefont
  {Wanjari}, \citenamefont {Gibb},\ and\ \citenamefont
  {Ashbaugh}}]{Ashbaugh&Gibb}%
  \BibitemOpen
  \bibfield  {author} {\bibinfo {author} {\bibfnamefont {P.~P.}\ \bibnamefont
  {Wanjari}}, \bibinfo {author} {\bibfnamefont {B.~C.}\ \bibnamefont {Gibb}}, \
  and\ \bibinfo {author} {\bibfnamefont {H.~S.}\ \bibnamefont {Ashbaugh}},\
  }\href@noop {} {\bibfield  {journal} {\bibinfo  {journal}
  {Annu.~Rev.~Phys.~Chem.}\ }\textbf {\bibinfo {volume} {67}},\ \bibinfo
  {pages} {617} (\bibinfo {year} {2016})}\BibitemShut {NoStop}%
\bibitem [{\citenamefont {Xue}\ \emph {et~al.}(2012)\citenamefont {Xue},
  \citenamefont {Yang}, \citenamefont {Chi}, \citenamefont {Zhang},\ and\
  \citenamefont {Huang}}]{Xue&Huang}%
  \BibitemOpen
  \bibfield  {author} {\bibinfo {author} {\bibfnamefont {M.}~\bibnamefont
  {Xue}}, \bibinfo {author} {\bibfnamefont {Y.}~\bibnamefont {Yang}}, \bibinfo
  {author} {\bibfnamefont {X.}~\bibnamefont {Chi}}, \bibinfo {author}
  {\bibfnamefont {Z.}~\bibnamefont {Zhang}}, \ and\ \bibinfo {author}
  {\bibfnamefont {F.}~\bibnamefont {Huang}},\ }\href@noop {} {\bibfield
  {journal} {\bibinfo  {journal} {Acc.~Chem.~Res.}\ }\textbf {\bibinfo {volume}
  {45}},\ \bibinfo {pages} {1294} (\bibinfo {year} {2012})}\BibitemShut
  {NoStop}%
\bibitem [{\citenamefont {Koshland}(1995)}]{key-lock1}%
  \BibitemOpen
  \bibfield  {author} {\bibinfo {author} {\bibfnamefont {D.~E.}\ \bibnamefont
  {Koshland}},\ }\href@noop {} {\bibfield  {journal} {\bibinfo  {journal}
  {Angew.~Chem.,~Int.~Ed.~Engl.}\ }\textbf {\bibinfo {volume} {33}},\ \bibinfo
  {pages} {2375} (\bibinfo {year} {1995})}\BibitemShut {NoStop}%
\bibitem [{\citenamefont {Hu}\ \emph {et~al.}(2009)\citenamefont {Hu},
  \citenamefont {Cheng}, \citenamefont {Wu}, \citenamefont {Zhao},\ and\
  \citenamefont {Xu}}]{host-guest1}%
  \BibitemOpen
  \bibfield  {author} {\bibinfo {author} {\bibfnamefont {J.}~\bibnamefont
  {Hu}}, \bibinfo {author} {\bibfnamefont {Y.}~\bibnamefont {Cheng}}, \bibinfo
  {author} {\bibfnamefont {Q.}~\bibnamefont {Wu}}, \bibinfo {author}
  {\bibfnamefont {L.}~\bibnamefont {Zhao}}, \ and\ \bibinfo {author}
  {\bibfnamefont {T.}~\bibnamefont {Xu}},\ }\href@noop {} {\bibfield  {journal}
  {\bibinfo  {journal} {J.~Phys.~Chem.~B}\ }\textbf {\bibinfo {volume} {113}},\
  \bibinfo {pages} {10650} (\bibinfo {year} {2009})}\BibitemShut {NoStop}%
\bibitem [{\citenamefont {Ball}(2008)}]{Ball:ChemRev}%
  \BibitemOpen
  \bibfield  {author} {\bibinfo {author} {\bibfnamefont {P.}~\bibnamefont
  {Ball}},\ }\href@noop {} {\bibfield  {journal} {\bibinfo  {journal}
  {Chem.~Rev.}\ }\textbf {\bibinfo {volume} {108}},\ \bibinfo {pages} {74}
  (\bibinfo {year} {2008})}\BibitemShut {NoStop}%
\bibitem [{\citenamefont {Young}\ \emph {et~al.}(2007)\citenamefont {Young},
  \citenamefont {Abel}, \citenamefont {Kim}, \citenamefont {Berne},\ and\
  \citenamefont {Friesner}}]{Berne&Friesner}%
  \BibitemOpen
  \bibfield  {author} {\bibinfo {author} {\bibfnamefont {T.}~\bibnamefont
  {Young}}, \bibinfo {author} {\bibfnamefont {R.}~\bibnamefont {Abel}},
  \bibinfo {author} {\bibfnamefont {B.}~\bibnamefont {Kim}}, \bibinfo {author}
  {\bibfnamefont {B.~J.}\ \bibnamefont {Berne}}, \ and\ \bibinfo {author}
  {\bibfnamefont {R.~A.}\ \bibnamefont {Friesner}},\ }\href@noop {} {\bibfield
  {journal} {\bibinfo  {journal} {Proc.~Natl.~Acad.~Sci.~(USA)}\ }\textbf
  {\bibinfo {volume} {104}},\ \bibinfo {pages} {808} (\bibinfo {year}
  {2007})}\BibitemShut {NoStop}%
\bibitem [{\citenamefont {Young}\ \emph {et~al.}(2010)\citenamefont {Young},
  \citenamefont {Hua}, \citenamefont {Huang}, \citenamefont {Abel},
  \citenamefont {Friesner},\ and\ \citenamefont {Berne}}]{Friesner&Berne}%
  \BibitemOpen
  \bibfield  {author} {\bibinfo {author} {\bibfnamefont {T.}~\bibnamefont
  {Young}}, \bibinfo {author} {\bibfnamefont {L.}~\bibnamefont {Hua}}, \bibinfo
  {author} {\bibfnamefont {X.}~\bibnamefont {Huang}}, \bibinfo {author}
  {\bibfnamefont {R.}~\bibnamefont {Abel}}, \bibinfo {author} {\bibfnamefont
  {R.~A.}\ \bibnamefont {Friesner}}, \ and\ \bibinfo {author} {\bibfnamefont
  {B.~J.}\ \bibnamefont {Berne}},\ }\href@noop {} {\bibfield  {journal}
  {\bibinfo  {journal} {Proteins}\ }\textbf {\bibinfo {volume} {78}},\ \bibinfo
  {pages} {1856} (\bibinfo {year} {2010})}\BibitemShut {NoStop}%
\bibitem [{\citenamefont {Nair}\ \emph {et~al.}(1991)\citenamefont {Nair},
  \citenamefont {Calderone}, \citenamefont {Christianson},\ and\ \citenamefont
  {Fierke}}]{Nair}%
  \BibitemOpen
  \bibfield  {author} {\bibinfo {author} {\bibfnamefont {S.~K.}\ \bibnamefont
  {Nair}}, \bibinfo {author} {\bibfnamefont {T.~L.}\ \bibnamefont {Calderone}},
  \bibinfo {author} {\bibfnamefont {D.~M.}\ \bibnamefont {Christianson}}, \
  and\ \bibinfo {author} {\bibfnamefont {C.~A.}\ \bibnamefont {Fierke}},\
  }\href@noop {} {\bibfield  {journal} {\bibinfo  {journal} {J.~Biol.~Chem.}\
  }\textbf {\bibinfo {volume} {266}},\ \bibinfo {pages} {17320} (\bibinfo
  {year} {1991})}\BibitemShut {NoStop}%
\bibitem [{\citenamefont {Carey}\ \emph {et~al.}(2000)\citenamefont {Carey},
  \citenamefont {Cheng},\ and\ \citenamefont {Rossky}}]{Rossky}%
  \BibitemOpen
  \bibfield  {author} {\bibinfo {author} {\bibfnamefont {C.}~\bibnamefont
  {Carey}}, \bibinfo {author} {\bibfnamefont {Y.-K.}\ \bibnamefont {Cheng}}, \
  and\ \bibinfo {author} {\bibfnamefont {P.~J.}\ \bibnamefont {Rossky}},\
  }\href@noop {} {\bibfield  {journal} {\bibinfo  {journal} {Chem.~Phys.}\
  }\textbf {\bibinfo {volume} {258}},\ \bibinfo {pages} {415} (\bibinfo {year}
  {2000})}\BibitemShut {NoStop}%
\bibitem [{\citenamefont {Setny}\ \emph {et~al.}(2009)\citenamefont {Setny},
  \citenamefont {Wang}, \citenamefont {Cheng}, \citenamefont {Li},
  \citenamefont {McCammon},\ and\ \citenamefont {Dzubiella}}]{Setny:PRL}%
  \BibitemOpen
  \bibfield  {author} {\bibinfo {author} {\bibfnamefont {P.}~\bibnamefont
  {Setny}}, \bibinfo {author} {\bibfnamefont {Z.}~\bibnamefont {Wang}},
  \bibinfo {author} {\bibfnamefont {L.-T.}\ \bibnamefont {Cheng}}, \bibinfo
  {author} {\bibfnamefont {B.}~\bibnamefont {Li}}, \bibinfo {author}
  {\bibfnamefont {J.~A.}\ \bibnamefont {McCammon}}, \ and\ \bibinfo {author}
  {\bibfnamefont {J.}~\bibnamefont {Dzubiella}},\ }\href@noop {} {\bibfield
  {journal} {\bibinfo  {journal} {Phys.~Rev.~Lett.}\ }\textbf {\bibinfo
  {volume} {103}},\ \bibinfo {pages} {187801} (\bibinfo {year}
  {2009})}\BibitemShut {NoStop}%
\bibitem [{\citenamefont {Setny}\ \emph {et~al.}(2010)\citenamefont {Setny},
  \citenamefont {Baron},\ and\ \citenamefont {McCammon}}]{Setny:JCTC:2010}%
  \BibitemOpen
  \bibfield  {author} {\bibinfo {author} {\bibfnamefont {P.}~\bibnamefont
  {Setny}}, \bibinfo {author} {\bibfnamefont {R.}~\bibnamefont {Baron}}, \ and\
  \bibinfo {author} {\bibfnamefont {J.~A.}\ \bibnamefont {McCammon}},\
  }\href@noop {} {\bibfield  {journal} {\bibinfo  {journal}
  {J.~Chem.~Theory~Comput.}\ }\textbf {\bibinfo {volume} {6}},\ \bibinfo
  {pages} {2866} (\bibinfo {year} {2010})}\BibitemShut {NoStop}%
\bibitem [{\citenamefont {Baron}\ \emph {et~al.}(2010)\citenamefont {Baron},
  \citenamefont {Setny},\ and\ \citenamefont {McCammon}}]{Setny:JACS}%
  \BibitemOpen
  \bibfield  {author} {\bibinfo {author} {\bibfnamefont {R.}~\bibnamefont
  {Baron}}, \bibinfo {author} {\bibfnamefont {P.}~\bibnamefont {Setny}}, \ and\
  \bibinfo {author} {\bibfnamefont {J.~A.}\ \bibnamefont {McCammon}},\
  }\href@noop {} {\bibfield  {journal} {\bibinfo  {journal}
  {J.~Am.~Chem.~Soc.}\ }\textbf {\bibinfo {volume} {132}},\ \bibinfo {pages}
  {12091} (\bibinfo {year} {2010})}\BibitemShut {NoStop}%
\bibitem [{\citenamefont {Pan}\ \emph {et~al.}(2013)\citenamefont {Pan},
  \citenamefont {Borhani}, \citenamefont {Dror},\ and\ \citenamefont
  {Shaw}}]{Pan&Shaw}%
  \BibitemOpen
  \bibfield  {author} {\bibinfo {author} {\bibfnamefont {A.~C.}\ \bibnamefont
  {Pan}}, \bibinfo {author} {\bibfnamefont {D.~W.}\ \bibnamefont {Borhani}},
  \bibinfo {author} {\bibfnamefont {R.~O.}\ \bibnamefont {Dror}}, \ and\
  \bibinfo {author} {\bibfnamefont {D.~E.}\ \bibnamefont {Shaw}},\ }\href@noop
  {} {\bibfield  {journal} {\bibinfo  {journal} {Drug~Dicov.~Today}\ }\textbf
  {\bibinfo {volume} {18}},\ \bibinfo {pages} {667} (\bibinfo {year}
  {2013})}\BibitemShut {NoStop}%
\bibitem [{\citenamefont {Setny}\ \emph {et~al.}(2013)\citenamefont {Setny},
  \citenamefont {Baron}, \citenamefont {Kekenes-Huskey}, \citenamefont
  {McCammon},\ and\ \citenamefont {Dzubiella}}]{Setny:PNAS}%
  \BibitemOpen
  \bibfield  {author} {\bibinfo {author} {\bibfnamefont {P.}~\bibnamefont
  {Setny}}, \bibinfo {author} {\bibfnamefont {R.}~\bibnamefont {Baron}},
  \bibinfo {author} {\bibfnamefont {P.}~\bibnamefont {Kekenes-Huskey}},
  \bibinfo {author} {\bibfnamefont {J.~A.}\ \bibnamefont {McCammon}}, \ and\
  \bibinfo {author} {\bibfnamefont {J.}~\bibnamefont {Dzubiella}},\ }\href@noop
  {} {\bibfield  {journal} {\bibinfo  {journal} {Proc.~Natl.~Acad.~Sci.~(USA)}\
  }\textbf {\bibinfo {volume} {110}},\ \bibinfo {pages} {1197} (\bibinfo {year}
  {2013})}\BibitemShut {NoStop}%
\bibitem [{\citenamefont {Setny}\ and\ \citenamefont
  {Geller}(2006)}]{Setny2006}%
  \BibitemOpen
  \bibfield  {author} {\bibinfo {author} {\bibfnamefont {P.}~\bibnamefont
  {Setny}}\ and\ \bibinfo {author} {\bibfnamefont {M.}~\bibnamefont {Geller}},\
  }\href@noop {} {\bibfield  {journal} {\bibinfo  {journal} {J.~Chem.~Phys.}\
  }\textbf {\bibinfo {volume} {125}},\ \bibinfo {pages} {14417} (\bibinfo
  {year} {2006})}\BibitemShut {NoStop}%
\bibitem [{\citenamefont {Setny}(2007)}]{Setny2007}%
  \BibitemOpen
  \bibfield  {author} {\bibinfo {author} {\bibfnamefont {P.}~\bibnamefont
  {Setny}},\ }\href@noop {} {\bibfield  {journal} {\bibinfo  {journal}
  {J.~Chem.~Phys.}\ }\textbf {\bibinfo {volume} {127}},\ \bibinfo {pages}
  {054505} (\bibinfo {year} {2007})}\BibitemShut {NoStop}%
\bibitem [{\citenamefont {Setny}(2008)}]{Setny2008}%
  \BibitemOpen
  \bibfield  {author} {\bibinfo {author} {\bibfnamefont {P.}~\bibnamefont
  {Setny}},\ }\href@noop {} {\bibfield  {journal} {\bibinfo  {journal}
  {J.~Chem.~Phys.}\ }\textbf {\bibinfo {volume} {128}},\ \bibinfo {pages}
  {125105} (\bibinfo {year} {2008})}\BibitemShut {NoStop}%
\bibitem [{\citenamefont {Mondal}\ \emph {et~al.}(2013)\citenamefont {Mondal},
  \citenamefont {Morrone},\ and\ \citenamefont {~}}]{Mondal:PNAS}%
  \BibitemOpen
  \bibfield  {author} {\bibinfo {author} {\bibfnamefont {J.}~\bibnamefont
  {Mondal}}, \bibinfo {author} {\bibfnamefont {J.~A.}\ \bibnamefont {Morrone}},
  \ and\ \bibinfo {author} {\bibfnamefont {B.~J.}\ \bibnamefont {~}},\
  }\href@noop {} {\bibfield  {journal} {\bibinfo  {journal}
  {Proc.~Natl.~Acad.~Sci.~(USA)}\ }\textbf {\bibinfo {volume} {110}},\ \bibinfo
  {pages} {13277} (\bibinfo {year} {2013})}\BibitemShut {NoStop}%
\bibitem [{\citenamefont {Wei{\ss}}\ \emph {et~al.}(2017)\citenamefont
  {Wei{\ss}}, \citenamefont {Setny},\ and\ \citenamefont
  {Dzubiella}}]{Weiss:JCTC}%
  \BibitemOpen
  \bibfield  {author} {\bibinfo {author} {\bibfnamefont {R.~G.}\ \bibnamefont
  {Wei{\ss}}}, \bibinfo {author} {\bibfnamefont {P.}~\bibnamefont {Setny}}, \
  and\ \bibinfo {author} {\bibfnamefont {J.}~\bibnamefont {Dzubiella}},\
  }\href@noop {} {\bibfield  {journal} {\bibinfo  {journal}
  {J.~Chem.~Theory~Comput.}\ }\textbf {\bibinfo {volume} {13}},\ \bibinfo
  {pages} {3012} (\bibinfo {year} {2017})}\BibitemShut {NoStop}%
\bibitem [{\citenamefont {Tiwary}\ \emph {et~al.}(2015)\citenamefont {Tiwary},
  \citenamefont {Mondal}, \citenamefont {Morrone},\ and\ \citenamefont
  {Berne}}]{Tiwary:PNAS}%
  \BibitemOpen
  \bibfield  {author} {\bibinfo {author} {\bibfnamefont {P.}~\bibnamefont
  {Tiwary}}, \bibinfo {author} {\bibfnamefont {J.}~\bibnamefont {Mondal}},
  \bibinfo {author} {\bibfnamefont {J.~A.}\ \bibnamefont {Morrone}}, \ and\
  \bibinfo {author} {\bibfnamefont {B.~J.}\ \bibnamefont {Berne}},\ }\href@noop
  {} {\bibfield  {journal} {\bibinfo  {journal} {Proc.~Natl.~Acad.~Sci.~(USA)}\
  }\textbf {\bibinfo {volume} {112}},\ \bibinfo {pages} {12015} (\bibinfo
  {year} {2015})}\BibitemShut {NoStop}%
\bibitem [{\citenamefont {Tiwary}\ and\ \citenamefont
  {Berne}(2016)}]{Tiwary:JCP}%
  \BibitemOpen
  \bibfield  {author} {\bibinfo {author} {\bibfnamefont {P.}~\bibnamefont
  {Tiwary}}\ and\ \bibinfo {author} {\bibfnamefont {B.~J.}\ \bibnamefont
  {Berne}},\ }\href@noop {} {\bibfield  {journal} {\bibinfo  {journal} {J.
  Chem. Phys.}\ }\textbf {\bibinfo {volume} {145}},\ \bibinfo {pages} {054113}
  (\bibinfo {year} {2016})}\BibitemShut {NoStop}%
\bibitem [{\citenamefont {Ludwig}\ \emph {et~al.}(2007)\citenamefont {Ludwig},
  \citenamefont {Michiels}, \citenamefont {Wu}, \citenamefont {Kavanagh},
  \citenamefont {Pilka}, \citenamefont {Jansson}, \citenamefont {Oppermann},\
  and\ \citenamefont {G{\"u}nther}}]{Ludwig2007}%
  \BibitemOpen
  \bibfield  {author} {\bibinfo {author} {\bibfnamefont {C.}~\bibnamefont
  {Ludwig}}, \bibinfo {author} {\bibfnamefont {P.~J.~A.}\ \bibnamefont
  {Michiels}}, \bibinfo {author} {\bibfnamefont {X.}~\bibnamefont {Wu}},
  \bibinfo {author} {\bibfnamefont {K.~L.}\ \bibnamefont {Kavanagh}}, \bibinfo
  {author} {\bibfnamefont {E.}~\bibnamefont {Pilka}}, \bibinfo {author}
  {\bibfnamefont {A.}~\bibnamefont {Jansson}}, \bibinfo {author} {\bibfnamefont
  {U.}~\bibnamefont {Oppermann}}, \ and\ \bibinfo {author} {\bibfnamefont
  {U.~L.}\ \bibnamefont {G{\"u}nther}},\ }\href@noop {} {\bibfield  {journal}
  {\bibinfo  {journal} {J. Med. Chem.}\ }\textbf {\bibinfo {volume} {51}},\
  \bibinfo {pages} {1} (\bibinfo {year} {2007})}\BibitemShut {NoStop}%
\bibitem [{\citenamefont {Wittmann}\ and\ \citenamefont
  {Strasser}(2017)}]{Wittmann2017}%
  \BibitemOpen
  \bibfield  {author} {\bibinfo {author} {\bibfnamefont {H.-J.}\ \bibnamefont
  {Wittmann}}\ and\ \bibinfo {author} {\bibfnamefont {A.}~\bibnamefont
  {Strasser}},\ }\href@noop {} {\bibfield  {journal} {\bibinfo  {journal}
  {Naunyn-Schmiedeberg's Arch. Pharmacol.}\ }\textbf {\bibinfo {volume}
  {390}},\ \bibinfo {pages} {595} (\bibinfo {year} {2017})}\BibitemShut
  {NoStop}%
\bibitem [{\citenamefont {Bruning}\ \emph {et~al.}(2010)\citenamefont
  {Bruning}, \citenamefont {Parent}, \citenamefont {Gil}, \citenamefont {Zhao},
  \citenamefont {Nowak}, \citenamefont {Pace}, \citenamefont {Smith},
  \citenamefont {Afonine}, \citenamefont {Adams}, \citenamefont
  {Katzenellenbogen},\ and\ \citenamefont {Nettles}}]{Bruning2010}%
  \BibitemOpen
  \bibfield  {author} {\bibinfo {author} {\bibfnamefont {J.~B.}\ \bibnamefont
  {Bruning}}, \bibinfo {author} {\bibfnamefont {A.~A.}\ \bibnamefont {Parent}},
  \bibinfo {author} {\bibfnamefont {G.}~\bibnamefont {Gil}}, \bibinfo {author}
  {\bibfnamefont {M.}~\bibnamefont {Zhao}}, \bibinfo {author} {\bibfnamefont
  {J.}~\bibnamefont {Nowak}}, \bibinfo {author} {\bibfnamefont {M.~C.}\
  \bibnamefont {Pace}}, \bibinfo {author} {\bibfnamefont {C.~L.}\ \bibnamefont
  {Smith}}, \bibinfo {author} {\bibfnamefont {P.~V.}\ \bibnamefont {Afonine}},
  \bibinfo {author} {\bibfnamefont {P.~D.}\ \bibnamefont {Adams}}, \bibinfo
  {author} {\bibfnamefont {J.~A.}\ \bibnamefont {Katzenellenbogen}}, \ and\
  \bibinfo {author} {\bibfnamefont {K.~W.}\ \bibnamefont {Nettles}},\
  }\href@noop {} {\bibfield  {journal} {\bibinfo  {journal} {Nat. Chem. Biol.}\
  }\textbf {\bibinfo {volume} {6}},\ \bibinfo {pages} {837} (\bibinfo {year}
  {2010})}\BibitemShut {NoStop}%
\bibitem [{\citenamefont {Takaoka}\ \emph {et~al.}(2010)\citenamefont
  {Takaoka}, \citenamefont {Ohta}, \citenamefont {Takeuchi}, \citenamefont
  {Miura}, \citenamefont {Matsuo}, \citenamefont {Sakaeda}, \citenamefont
  {Sugano},\ and\ \citenamefont {Nishio}}]{Takaoka2010}%
  \BibitemOpen
  \bibfield  {author} {\bibinfo {author} {\bibfnamefont {Y.}~\bibnamefont
  {Takaoka}}, \bibinfo {author} {\bibfnamefont {M.}~\bibnamefont {Ohta}},
  \bibinfo {author} {\bibfnamefont {A.}~\bibnamefont {Takeuchi}}, \bibinfo
  {author} {\bibfnamefont {K.}~\bibnamefont {Miura}}, \bibinfo {author}
  {\bibfnamefont {M.}~\bibnamefont {Matsuo}}, \bibinfo {author} {\bibfnamefont
  {T.}~\bibnamefont {Sakaeda}}, \bibinfo {author} {\bibfnamefont
  {A.}~\bibnamefont {Sugano}}, \ and\ \bibinfo {author} {\bibfnamefont
  {H.}~\bibnamefont {Nishio}},\ }\href@noop {} {\bibfield  {journal} {\bibinfo
  {journal} {J. Biochem.}\ }\textbf {\bibinfo {volume} {148}},\ \bibinfo
  {pages} {25} (\bibinfo {year} {2010})}\BibitemShut {NoStop}%
\bibitem [{\citenamefont {Benkaidali}\ \emph {et~al.}(2013)\citenamefont
  {Benkaidali}, \citenamefont {Andr{\'e}}, \citenamefont {Maouche},
  \citenamefont {Siregar}, \citenamefont {Benyettou}, \citenamefont {Maurel},\
  and\ \citenamefont {Petitjean}}]{Benkaidali2013}%
  \BibitemOpen
  \bibfield  {author} {\bibinfo {author} {\bibfnamefont {L.}~\bibnamefont
  {Benkaidali}}, \bibinfo {author} {\bibfnamefont {F.}~\bibnamefont
  {Andr{\'e}}}, \bibinfo {author} {\bibfnamefont {B.}~\bibnamefont {Maouche}},
  \bibinfo {author} {\bibfnamefont {P.}~\bibnamefont {Siregar}}, \bibinfo
  {author} {\bibfnamefont {M.}~\bibnamefont {Benyettou}}, \bibinfo {author}
  {\bibfnamefont {F.}~\bibnamefont {Maurel}}, \ and\ \bibinfo {author}
  {\bibfnamefont {M.}~\bibnamefont {Petitjean}},\ }\href@noop {} {\bibfield
  {journal} {\bibinfo  {journal} {Bioinformatics}\ }\textbf {\bibinfo {volume}
  {30}},\ \bibinfo {pages} {792} (\bibinfo {year} {2013})}\BibitemShut
  {NoStop}%
\bibitem [{\citenamefont {Kinoshito}(2004)}]{Kinoshito}%
  \BibitemOpen
  \bibfield  {author} {\bibinfo {author} {\bibfnamefont {M.}~\bibnamefont
  {Kinoshito}},\ }\href@noop {} {\bibfield  {journal} {\bibinfo  {journal}
  {Chem.~Phys.~Lett.}\ }\textbf {\bibinfo {volume} {387}},\ \bibinfo {pages}
  {47} (\bibinfo {year} {2004})}\BibitemShut {NoStop}%
\bibitem [{\citenamefont {Roth}\ \emph {et~al.}(2002)\citenamefont {Roth},
  \citenamefont {van Roij}, \citenamefont {Andrienko}, \citenamefont {Mecke},\
  and\ \citenamefont {Dietrich}}]{Roth}%
  \BibitemOpen
  \bibfield  {author} {\bibinfo {author} {\bibfnamefont {R.}~\bibnamefont
  {Roth}}, \bibinfo {author} {\bibfnamefont {R.}~\bibnamefont {van Roij}},
  \bibinfo {author} {\bibfnamefont {D.}~\bibnamefont {Andrienko}}, \bibinfo
  {author} {\bibfnamefont {K.~R.}\ \bibnamefont {Mecke}}, \ and\ \bibinfo
  {author} {\bibfnamefont {S.}~\bibnamefont {Dietrich}},\ }\href@noop {}
  {\bibfield  {journal} {\bibinfo  {journal} {Phys.~Rev.~Lett.}\ }\textbf
  {\bibinfo {volume} {89}} (\bibinfo {year} {2002})}\BibitemShut {NoStop}%
\bibitem [{\citenamefont {K{\"o}nig}\ \emph {et~al.}(2008)\citenamefont
  {K{\"o}nig}, \citenamefont {Roth},\ and\ \citenamefont {Dietrich}}]{König}%
  \BibitemOpen
  \bibfield  {author} {\bibinfo {author} {\bibfnamefont {P.-M.}\ \bibnamefont
  {K{\"o}nig}}, \bibinfo {author} {\bibfnamefont {R.}~\bibnamefont {Roth}}, \
  and\ \bibinfo {author} {\bibfnamefont {S.}~\bibnamefont {Dietrich}},\
  }\href@noop {} {\bibfield  {journal} {\bibinfo  {journal} {Europhys.~Lett.}\
  }\textbf {\bibinfo {volume} {84}} (\bibinfo {year} {2008})}\BibitemShut
  {NoStop}%
\bibitem [{\citenamefont {D{\l{}}ugosz}\ and\ \citenamefont
  {Antosiewicz}(2016)}]{Dlugosz}%
  \BibitemOpen
  \bibfield  {author} {\bibinfo {author} {\bibfnamefont {M.}~\bibnamefont
  {D{\l{}}ugosz}}\ and\ \bibinfo {author} {\bibfnamefont {J.~M.}\ \bibnamefont
  {Antosiewicz}},\ }\href@noop {} {\bibfield  {journal} {\bibinfo  {journal}
  {J.~Phys.~Chem.~B}\ }\textbf {\bibinfo {volume} {120}},\ \bibinfo {pages}
  {7114} (\bibinfo {year} {2016})}\BibitemShut {NoStop}%
\bibitem [{\citenamefont {Wei{\ss}}\ \emph {et~al.}(2016)\citenamefont
  {Wei{\ss}}, \citenamefont {Setny},\ and\ \citenamefont
  {Dzubiella}}]{Weiss:JPCB}%
  \BibitemOpen
  \bibfield  {author} {\bibinfo {author} {\bibfnamefont {R.~G.}\ \bibnamefont
  {Wei{\ss}}}, \bibinfo {author} {\bibfnamefont {P.}~\bibnamefont {Setny}}, \
  and\ \bibinfo {author} {\bibfnamefont {J.}~\bibnamefont {Dzubiella}},\
  }\href@noop {} {\bibfield  {journal} {\bibinfo  {journal} {J.~Phys.~Chem.~B}\
  }\textbf {\bibinfo {volume} {120}},\ \bibinfo {pages} {8127} (\bibinfo {year}
  {2016})}\BibitemShut {NoStop}%
\bibitem [{\citenamefont {Hinczewski}\ \emph {et~al.}(2010)\citenamefont
  {Hinczewski}, \citenamefont {von Hansen}, \citenamefont {Dzubiella},\ and\
  \citenamefont {Netz}}]{Hinczewski&Netz}%
  \BibitemOpen
  \bibfield  {author} {\bibinfo {author} {\bibfnamefont {M.}~\bibnamefont
  {Hinczewski}}, \bibinfo {author} {\bibfnamefont {Y.}~\bibnamefont {von
  Hansen}}, \bibinfo {author} {\bibfnamefont {J.}~\bibnamefont {Dzubiella}}, \
  and\ \bibinfo {author} {\bibfnamefont {R.~R.}\ \bibnamefont {Netz}},\
  }\href@noop {} {\bibfield  {journal} {\bibinfo  {journal} {J. Chem. Phys.}\
  }\textbf {\bibinfo {volume} {132}},\ \bibinfo {pages} {245103} (\bibinfo
  {year} {2010})}\BibitemShut {NoStop}%
\bibitem [{\citenamefont {Jorgensen}\ \emph {et~al.}(1996)\citenamefont
  {Jorgensen}, \citenamefont {Maxwell},\ and\ \citenamefont
  {Tirado-Rives}}]{Jorgensen1996OPLS-AA}%
  \BibitemOpen
  \bibfield  {author} {\bibinfo {author} {\bibfnamefont {W.~L.}\ \bibnamefont
  {Jorgensen}}, \bibinfo {author} {\bibfnamefont {D.~S.}\ \bibnamefont
  {Maxwell}}, \ and\ \bibinfo {author} {\bibfnamefont {J.}~\bibnamefont
  {Tirado-Rives}},\ }\href@noop {} {\bibfield  {journal} {\bibinfo  {journal}
  {J. Am. Chem. Soc}\ }\textbf {\bibinfo {volume} {118}},\ \bibinfo {pages}
  {11225} (\bibinfo {year} {1996})}\BibitemShut {NoStop}%
\bibitem [{\citenamefont {Jorgensen}\ \emph {et~al.}(1983)\citenamefont
  {Jorgensen}, \citenamefont {Chandrasekhar}, \citenamefont {Madura},
  \citenamefont {Impey},\ and\ \citenamefont {Klein}}]{Jorgensen1983TIP4P}%
  \BibitemOpen
  \bibfield  {author} {\bibinfo {author} {\bibfnamefont {W.~L.}\ \bibnamefont
  {Jorgensen}}, \bibinfo {author} {\bibfnamefont {J.}~\bibnamefont
  {Chandrasekhar}}, \bibinfo {author} {\bibfnamefont {J.~D.}\ \bibnamefont
  {Madura}}, \bibinfo {author} {\bibfnamefont {R.~W.}\ \bibnamefont {Impey}}, \
  and\ \bibinfo {author} {\bibfnamefont {M.~L.}\ \bibnamefont {Klein}},\
  }\href@noop {} {\bibfield  {journal} {\bibinfo  {journal} {J. Chem. Phys.}\
  }\textbf {\bibinfo {volume} {79}},\ \bibinfo {pages} {926} (\bibinfo {year}
  {1983})}\BibitemShut {NoStop}%
\bibitem [{\citenamefont {Jorgensen}\ and\ \citenamefont
  {Madura}(1985)}]{Jorgensen1985TIP4P}%
  \BibitemOpen
  \bibfield  {author} {\bibinfo {author} {\bibfnamefont {W.~L.}\ \bibnamefont
  {Jorgensen}}\ and\ \bibinfo {author} {\bibfnamefont {J.~D.}\ \bibnamefont
  {Madura}},\ }\href@noop {} {\bibfield  {journal} {\bibinfo  {journal} {Mol.
  Phys.}\ }\textbf {\bibinfo {volume} {56}},\ \bibinfo {pages} {1381} (\bibinfo
  {year} {1985})}\BibitemShut {NoStop}%
\bibitem [{\citenamefont {Kumar}\ \emph {et~al.}(1992)\citenamefont {Kumar},
  \citenamefont {Bouzida}, \citenamefont {Swendsen}, \citenamefont {Kollman},\
  and\ \citenamefont {Rosenberg}}]{WHAM}%
  \BibitemOpen
  \bibfield  {author} {\bibinfo {author} {\bibfnamefont {S.}~\bibnamefont
  {Kumar}}, \bibinfo {author} {\bibfnamefont {D.}~\bibnamefont {Bouzida}},
  \bibinfo {author} {\bibfnamefont {R.~H.}\ \bibnamefont {Swendsen}}, \bibinfo
  {author} {\bibfnamefont {P.~A.}\ \bibnamefont {Kollman}}, \ and\ \bibinfo
  {author} {\bibfnamefont {J.~M.}\ \bibnamefont {Rosenberg}},\ }\href@noop {}
  {\bibfield  {journal} {\bibinfo  {journal} {J.~Comput.~Chem.}\ }\textbf
  {\bibinfo {volume} {13}},\ \bibinfo {pages} {1011} (\bibinfo {year}
  {1992})}\BibitemShut {NoStop}%
\bibitem [{\citenamefont {Zhu}\ and\ \citenamefont
  {Hummer}(2012)}]{HummerWHAM}%
  \BibitemOpen
  \bibfield  {author} {\bibinfo {author} {\bibfnamefont {F.}~\bibnamefont
  {Zhu}}\ and\ \bibinfo {author} {\bibfnamefont {G.}~\bibnamefont {Hummer}},\
  }\href@noop {} {\bibfield  {journal} {\bibinfo  {journal} {J.~Comput.~Chem.}\
  }\textbf {\bibinfo {volume} {33}},\ \bibinfo {pages} {453} (\bibinfo {year}
  {2012})}\BibitemShut {NoStop}%
\bibitem [{\citenamefont {Weiss}(1967)}]{FirstPassageWeiss}%
  \BibitemOpen
  \bibfield  {author} {\bibinfo {author} {\bibfnamefont {G.~H.}\ \bibnamefont
  {Weiss}},\ }\href@noop {} {\bibfield  {journal} {\bibinfo  {journal} {Adv.
  Chem. Phys.}\ }\textbf {\bibinfo {volume} {13}},\ \bibinfo {pages} {1}
  (\bibinfo {year} {1967})}\BibitemShut {NoStop}%
\bibitem [{\citenamefont {Siegert}(1951)}]{Siegert}%
  \BibitemOpen
  \bibfield  {author} {\bibinfo {author} {\bibfnamefont {A.~J.~F.}\
  \bibnamefont {Siegert}},\ }\href@noop {} {\bibfield  {journal} {\bibinfo
  {journal} {Phys. Rev.}\ }\textbf {\bibinfo {volume} {81}},\ \bibinfo {pages}
  {617} (\bibinfo {year} {1951})}\BibitemShut {NoStop}%
\bibitem [{\citenamefont {Morrone}\ \emph {et~al.}(2012)\citenamefont
  {Morrone}, \citenamefont {Li},\ and\ \citenamefont {Berne}}]{MorroneLiBerne}%
  \BibitemOpen
  \bibfield  {author} {\bibinfo {author} {\bibfnamefont {J.~A.}\ \bibnamefont
  {Morrone}}, \bibinfo {author} {\bibfnamefont {J.}~\bibnamefont {Li}}, \ and\
  \bibinfo {author} {\bibfnamefont {B.~J.}\ \bibnamefont {Berne}},\ }\href@noop
  {} {\bibfield  {journal} {\bibinfo  {journal} {J.~Phys.~Chem.~B}\ }\textbf
  {\bibinfo {volume} {116}},\ \bibinfo {pages} {378} (\bibinfo {year}
  {2012})}\BibitemShut {NoStop}%
\bibitem [{\citenamefont {Li}\ \emph {et~al.}(2012)\citenamefont {Li},
  \citenamefont {Morrone},\ and\ \citenamefont {Berne}}]{LiMorroneBerne}%
  \BibitemOpen
  \bibfield  {author} {\bibinfo {author} {\bibfnamefont {J.}~\bibnamefont
  {Li}}, \bibinfo {author} {\bibfnamefont {J.~A.}\ \bibnamefont {Morrone}}, \
  and\ \bibinfo {author} {\bibfnamefont {B.~J.}\ \bibnamefont {Berne}},\
  }\href@noop {} {\bibfield  {journal} {\bibinfo  {journal} {J.~Phys.~Chem.~B}\
  }\textbf {\bibinfo {volume} {116}},\ \bibinfo {pages} {11537} (\bibinfo
  {year} {2012})}\BibitemShut {NoStop}%
\end{thebibliography}
\end{document}


\bibliographystyle{apsrev4-1} 

\title{Supporting information for: Affinity, kinetics, and pathways of anisotropic ligands binding to hydrophobic model pockets}

\author{R. Gregor Wei{\ss}}
\thanks{To whom correspondence should be addressed. E-mail: gregor.weiss@physik.hu-berlin.de or joachim.dzubiella@helmholtz-berlin.de}
\affiliation{Institut f{\"u}r Physik, Humboldt-Universit{\"a}t zu Berlin, Newtonstr.~15, D-12489 Berlin, Germany}
\affiliation{Institut f{\"u}r Weiche Materie und Funktionale Materialen, Helmholtz-Zentrum Berlin, Hahn-Meitner-Platz 1, D-14109 Berlin, Germany}
\author{Richard Chudoba}                                                                                                             
\affiliation{Institut f{\"u}r Physik, Humboldt-Universit{\"a}t zu Berlin, Newtonstr.~15, D-12489 Berlin, Germany}                    
\affiliation{Institut f{\"u}r Weiche Materie und Funktionale Materialen, Helmholtz-Zentrum Berlin, Hahn-Meitner-Platz 1, D-14109 Berlin, Germany}
\author{Piotr Setny}
\affiliation{Centre of New Technologies, University of Warsaw, 00-927 Warsaw, Poland}
\author{Joachim Dzubiella}
\thanks{To whom correspondence should be addressed. E-mail: gregor.weiss@physik.hu-berlin.de or joachim.dzubiella@helmholtz-berlin.de}
\affiliation{Institut f{\"u}r Physik, Humboldt-Universit{\"a}t zu Berlin, Newtonstr.~15, D-12489 Berlin, Germany}
\affiliation{Institut f{\"u}r Weiche Materie und Funktionale Materialen, Helmholtz-Zentrum Berlin, Hahn-Meitner-Platz 1, D-14109 Berlin, Germany}

\maketitle

\section{Bulk friction constants of the aromatic ligands}

\begin{table}[b]
\caption{The different aromatic compounds have different bulk friction values 
$\beta\xi_\infty$. In general, the more elongated/bigger ligands have larger friction
constants.}
\begin{center}
\begin{tabular}{l | c}
 & $\beta\xi_\infty$ [ns~nm$^\mrm{-2}$] \\
\hline
ethylbenzene  	& $0.91$ \\
toluene       	& $0.84$ \\
benzene       	& $0.74$ \\
phenol        	& $0.82$ \\
benzyl alcohol 	& $0.87$ \\
sphere					& $0.40$ \\
\end{tabular}
\end{center}
\label{tab1}
\end{table}

\noindent
Here we briefly summarize the bulk friction constants of aromatic compounds. In general we can probe the bulk friction in
umbrella windows that are far away from the binding site by the position autocorrelation~\cite{HummerFriction} 
\begin{equation}
\beta \xi_\infty = \frac{\int_0^\infty \langle \delta z(t) \delta z (0) \rangle \dd t}{\langle \delta z^2 \rangle^2}
\end{equation}
where $\langle \delta z(t) \delta z (0) \rangle$ is the position auto-correlation function with $\delta z(t) =
z(t)-\langle z \rangle$. For the spherical ligand we know that the bulk friction is around
$\beta\xi_\infty=0.4~\mrm{ns}~\mrm{nm}^{-2}$
from our previous work~\cite{Weiss:JCTC}. According to Stokes friction the bulk friction should increase with ligand size which
holds for our aromatic compounds. Table~\ref{tab1} lists the bulk friction for ethylbenzene, toluene, benzene, phenol, benzyl
alcohol, and the spherical ligand. Throughout the paper we used these values to normalize the mean binding time and kinetic
profiles. This enabled us to compare the effective differences between our various ligands.

\section{MFPT of the aromatic ligands}

\begin{figure*}[t]
\centering \includegraphics{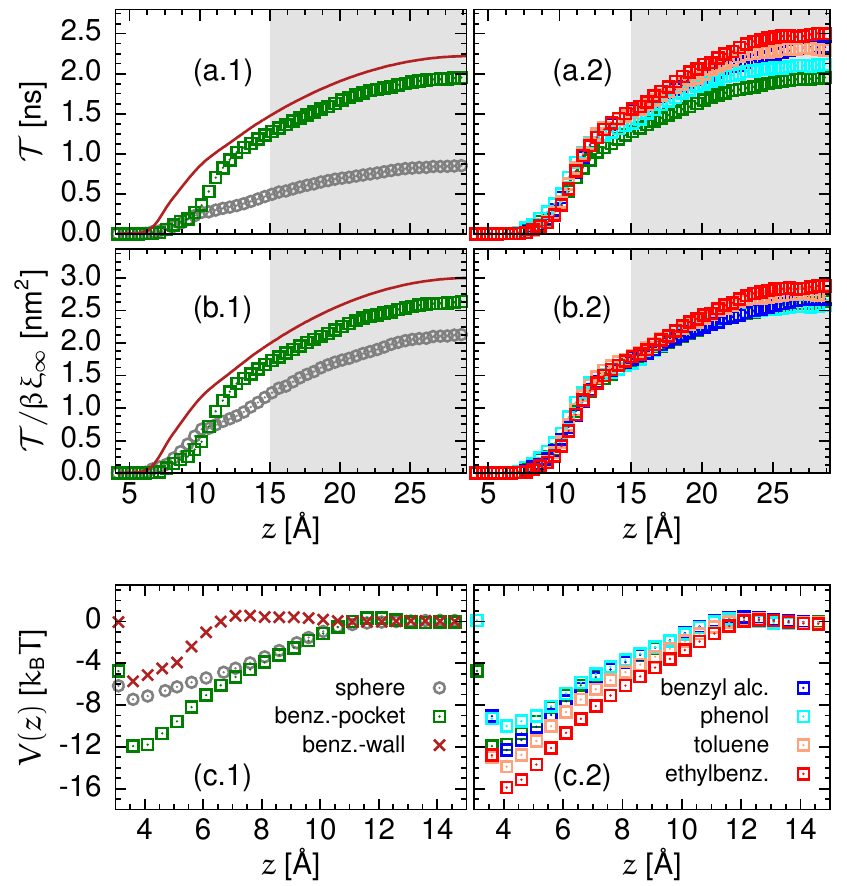}
\caption[MFPT for aromatic ligands.]{Panel (a.1) shows the MFPT curves $\mT$ for benzene binding to the wall and to the pocket as
well as binding of the spherical ligand to the pocket. Panel (a.2) compares the MFPT curves for benzyl alcohol, phenol, benzene,
toluene, and ethylbenzene. In agreement with Table~\ref{tab1} the bigger ligands with larger friction values bind overall
slower. In panels (b.1) and (b.2) we plot the MFPT results which are rescaled by the respective bulk friction value such that
$\mT/\beta\xi_\infty$. Panels (c.1) and (c.2) plot all PMFs for all pairs of ligands and binding sites.}
\label{figD1}
\end{figure*}

In Fig.~\ref{figD1} we plot the MFPT curves for all aromatic and the spherical ligand. The upper
panels (a) and (b) show the raw data $\mT(z,z_f)$ which makes evident that, e.g., ethylbenzene
binds slowest, however, this is a trivial result because the bulk friction constant of
ethylbenzene is the largest. If we rescale the MFPT to the respective bulk friction constants
the curves for benzene, phenol and benzyl alcohol mostly coincide in panel (d). Hence the
overall rescaled kinetics are in good agreement. This is consistent with the overlapping kinetic
barriers for these ligands in Fig.~5 in the main text. The rescaled kinetics
$\mT/\beta\xi_\infty$ of ethylbenzene and toluene are, however, slightly decelerated in
comparison to the benzene binding times. Thus here the deviations of the kinetic barrier in
Fig.~5 qualitatively agrees with the normalized MFPT curves here. Moreover benzene clearly binds
slower than the spherical ligand because its kinetic barrier is shifted further away from the
pocket.

%
